\documentclass[fleqn,usenatbib]{mnras}

\usepackage{newtxtext,newtxmath}

\usepackage[T1]{fontenc}
\usepackage{ae,aecompl}
\usepackage[flushleft]{threeparttable}

\DeclareRobustCommand{\VAN}[3]{#2}
\let\VANthebibliography\thebibliography
\def\thebibliography{\DeclareRobustCommand{\VAN}[3]{##3}\VANthebibliography}

\usepackage{graphicx}	% Including figure files
\usepackage{amsmath}	% Advanced maths commands
\usepackage{amssymb}	% Extra maths symbols
\usepackage[dvipsnames]{xcolor}
\usepackage{soul}

\defcitealias{Savorgnan2016A}{SG16}
\defcitealias{Davis2019}{D19}
\defcitealias{Sahu2019A}{S19}
\defcitealias{Hon2022}{H22}

\title[Spheroid size-mass relation]{The size--mass and other structural parameter ($n, \mu_z, R_z$) relations for local bulges/spheroids from multicomponent decompositions}

\author[Hon et al.]{
Dexter S. -H. Hon,$^{1}$\thanks{E-mail:~\href{mailto:dex-hon-sci@outlook.com}{dex-hon-sci@outlook.com}, \href{mailto:shon@swin.edu.au}{shon@swin.edu.au}}
Alister W. Graham,$^{1}$
and Nandini Sahu$^{1}$
\\
$^{1}$Centre for Astrophysics and Supercomputing, Swinburne University of Technology, Hawthorn, Victoria 3122, Australia}

\date{Last update 6 Dec 2022}

\pubyear{2022}

\begin{document}
\label{firstpage}
\pagerange{\pageref{firstpage}--\pageref{lastpage}}
\maketitle

\begin{abstract}
%point one: after careful decomposition, the size mass relation exhibit a curved feature

%Context-> Goal and hope -> methods -> result and discussion

%Spheroids/Bulges in the centre of galaxies are known to be an ancient structure that contain old stars.

%E and Sph are the same population

%With the advancement of multi-component decomposition, the structure of embedded spheroids 
We analyse the bulge/spheroid size--(stellar mass), $R_\mathrm{e,Sph}$--$M_\mathrm{*,Sph}$, relation and spheroid structural parameters for 202 local (predominantly $\lesssim 110~\rm Mpc$) galaxies spanning $ M_*\sim 3\times10^{9}$--$10^{12}~\rm M_{\odot}$ and $ 0.1 \lesssim R_\mathrm{e, Sph}\lesssim32~\rm kpc$ from multicomponent decomposition.
The correlations between the spheroid S\'ersic index ($n_\mathrm{Sph}$), central surface brightness ($\mu_\mathrm{0, Sph}$), effective half-light radius ($R_\mathrm{e, Sph}$), absolute magnitude ($\mathfrak{M}_\mathrm{Sph}$) and stellar mass ($M_\mathrm{*,Sph}$) are explored. 
We also investigate the consequences of using different scale radii, $R_{z,\rm Sph}$, encapsulating a different fraction ($z$, from 0 to 1) of the total spheroid luminosity.
The correlation strengths for projected mass densities, $\Sigma_z$ and $\langle \Sigma \rangle_z$, vary significantly with the choice of $z$.
Spheroid size ($R_\mathrm{z, \rm Sph}$) and mass ($M_\mathrm{*,Sph}$) are strongly correlated for all light fractions $z$.
We find: $\log(R_\mathrm{e,Sph}/\rm kpc) =   0.88\log(M_\mathrm{*,Sph}/\rm M_{\odot})-9.15$ with a small scatter of $\Delta_{rms} = 0.24~\rm dex$ in the $\log(R_\mathrm{e,Sph})$ direction. 
This result is discussed relative to the \textit{curved} size--mass relation for early-type galaxies due to their discs yielding larger galaxy radii at lower masses.
Moreover, the slope of our spheroid size--mass relation is a factor of $\sim3$, steeper than reported bulge size--mass relations, and with bulge sizes at $M_{\rm *,sph}\sim3\times10^9$~M$_\odot$ which are 2 to 3 times smaller. 
Our spheroid size--mass relation present no significant flattening in slope in the low-mass end ($M_{\rm *,sph}\sim10^9$--$10^{10}\rm~M_{\odot}$). 
Instead of treating galaxies as single entities, future theoretical and evolutionary models should also attempt to recreate the strong scaling relations of specific galactic components.
Additional scaling relations, such as $\log(n_\mathrm{Sph})$--$\log(M_\mathrm{*,Sph})$, $\log(\Sigma_\mathrm{0, Sph})$--$\log(n_\mathrm{Sph})$, and $\log(n_\mathrm{Sph})$--$\log(R_\mathrm{e,Sph})$, are also presented.
Finally, we show that the local spheroids align well with the size-mass distribution of quiescent galaxies at $z\sim1.25$--$2.25$. 
In essence, local spheroids and high-$z$ quiescent galaxies appear structurally similar, likely dictated by the virial theorem.

\end{abstract}

%%%%%%%%%%%%%%%%%%%%%%%%%%%%%%%%%%%%%%%%%%%%%%%%%%

%%%%%%%%%%%%%%%%% BODY OF PAPER %%%%%%%%%%%%%%%%%%

\section{Introduction} 
\label{sec:intro}

Some of the complexity in galaxy morphology arises from the different developmental processes of the ellipsoidal/spherical bulge and flat disc.
Their diametrically different nature leads to the support for a two phase formation scenario \citep[e.g.,][]{2013pss6.book...91G, Driver2013}, where the bulges were formed via a rapid, hot-mode process (early collapse) at high-redshifts \citep[e.g.,][]{Naab2009,Hopkins2009,Bezanson2009,Trujillo2011} and the discs are subsequently built through mergers and accretion \citep[][and reference therein]{Larson1976,Tinsley1978,Graham2015}. 
In this picture, massive bulges formed first ($z \gtrsim 2.0$) and the disc later, a.k.a., an inside-out evolution. 
Less massive galaxies may have experienced a delayed, more gradual evolution in a down-sizing scenario \citep[e.g.][]{Graham2017}. 
The third phase of evolution will be the major merging of these galaxies to produce elliptical (E) galaxies or S0 galaxies if the net angular momentum is not sufficiently cancelled.

While the bulges of some of today's massive galaxies appear to be the descendants of the high-$z$ compact massive galaxies \citep[such as NGC~3311, see][]{Barbosa2021}, some lower mass bulges may also have existed as smaller "red gems" in the distant Universe. 
Indeed, the stellar population in the inner part of galaxies seems to be significantly older than the outer region \citep[][]{Proctor2002, Moorthy2006, Thomas2006, Jablonka2007, MacArthur2008, Saracco2009}, with the work by \citet{MacArthur2009} revealing that the bulk of the stars in spiral galaxy bulges are old.

Early-type galaxies (ETGs) are known to follow a curved size-mass relation, stretching from the ellipitcal (E) galaxies at the high-mass end \citep[][]{Caon1993,D'Onofrio1994, 2003MNRAS.343..978S, Cappellari2011a, Baldry2012, Lange2015, Lange2016, Morishita2017, Nedkova2021, Noordeh2021} to the dwarf ETGs (dETGs) at the low-mass end \citep{Binggeli1998, Graham2006A, Forbes2008, 2013pss6.book...91G}.
Due to the prevalence of discs in ETGs \citep[][and reference therein]{Scott2015}, this curved relation is a product of the disc size and mass, which dominate the ETGs at the low-mas end.
In the 70s, when the galaxies' size was measured using isophotal radius \citep[e.g.,][]{Heidmann1969,Holmberg1969,Oemler1976,Strom1978}, the galaxies exhibited a linear size-mass relation in log-log space (i.e., the size--mass relation follows a simple power law)\footnote{Evidence of a linear isophotal size--mass relation are also seen in later works \citep[e.g.,][]{Forbes2008,van_den_Bergh2008,Nair2011}.}.
However, these works primarily focus on the size-mass relation for galaxies as a whole instead of their substructures.
Additional insight can be drawn if a similar study is conducted on the \textit{bulge/spheroids}. 
A bulge/bar/disc decomposition is required to explore the size--mass relation of ETG bulges, as opposed to the size--mass relation of the galaxies themselves.
This enables greater insight into the formation physics shaping the components of galaxies. 
Before high-resolution observations of distant galaxies, such as those promised by MAVIS (optical, under development, \citealt{e31b3f4abec045d38febe6f157a1e2e6,Rigaut2020MAVISconcept}) and VLT's PIONIER (1.6\,$\micron$, \citealt{Le_Bouquin2011PIONIER}), and GRAVITY (2.0--2.4\,$\micron$ imager, \citealt{Gillessen2010GRAVITY,Eisenhauer2011GRAVITY}) are available, local spheroids are a great place to probe the Universe's past.
Important questions can be asked in regard to bulge formation.
For instance, with the low merger rate at $ z < 0.7$  \citep[][]{De_Propris2005A,De_Propris2007A,De_Propris2010A}, how are spheroids formed so efficiently 10--12 Gyr ago?
Is the continuity in the spheroid size--mass relation, from E galaxies to spheroids embedded in S0, S, and dETGs, a coincidence, or are they the products of the same formation physics but simply differ in scale?

For decades, the standard method to break down galaxy structure was to fit either an $R^{1/4}$ \citep[][]{de_Vaucouleurs1948} or an $R^{1/n}$ \citep[][]{Sersic1968} model to describe the spheroid, plus an exponential function to describe the disc \citep[e.g.,][etc.]{Andredakis1995,Seigar1998,Iodice1999,Khosroshahi2000A,D'Onofrio2001,Graham2001A,Mollenhoff2001,Simard2002,Allen2006,Fisher2010,Simard2011}.
However, due to the complex substructures in galaxies, a simple Bulge+Disc decomposition is inappropriate for some.
Consequently, past size--mass relations for bulges are questionable due to the influence of, for example, the bar, which can inflate both the size and the mass of the presumed `bulge' component. 
This can muddy the waters when attempting to identify classical bulges versus `pseudo-bulges'.
Given the need for more detailed breakdowns of galaxy structures, many \citep[e.g.,][]{Martin1995,Prieto1997,Aguerri1998,Graham2003,Laurikainen2005,Gadotti2009,Laurikainen2010,Vika2012,Lasker2014,Savorgnan2016A,Davis2019,Sahu2019A} developed sophisticated, manual (i.e., not blind automated) multicomponent decompositions in an attempt to capture these substructures. 
The benefit of such, interestingly, is that now the spheroids can be better qualified because the biasing substructures have been accounted for.

While spheroids can grow, the entropy of such pressure-supported systems makes them hard to erase.
As such, they are expected to have longevity which bars and spiral arms will not.
Here, we focus on the spheroids, leaving the bar or disc scaling relation for future investigation.
We take advantage of multicomponent decompositions to examine the \textit{bulge/spheroid} size-mass ($R_\mathrm{e,Sph}$--$M_\mathrm{*,Sph}$) relation and some of their structural parameters: central surface brightness ($\mu_\mathrm{0,Sph}$) and S\'ersic index ($n_\mathrm{Sph}$).

In Section~\ref{sec:data_analysis}, we briefly describe our data sources (Section~\ref{sec:data}), the surface brightness profile extraction (Section~\ref{sec:SBanalysis}), and the models used to measure the spheroids (Section~\ref{sec:sph_mod}).
In Section~\ref{sec:Result}, we first discuss 
the spheroid structural parameters and their correlations with each others (Section~\ref{sec:Struc_para} and \ref{sec:Corr_among_para}).
We subsequently explore the changes among these parameters when using scale radii enclosing different fractions of the spheroid light (Section~\ref{sec:scale_radius}).
Subsequently, we converted the spheroid absolute magnitude into stellar mass (Section~\ref{sec:mag_to_mass}) and studied how the spheroid mass related scaling relations varies when using different scale radii (Section~\ref{sec:sigma0}, \ref{sec:average_sigma0}, and \ref{sec:corr_trend}).
The spheroid size-mass relation (Section~\ref{sec:size_mass}) and other scaling relations (Section~\ref{sec:additional_relations}) are also presented. 
Section~\ref{sec:Discussion} compares the spheroid size-mass relation to relevant works in the literature and discusses what it informs us on the formation history of the spheroids.

\section{Data and Analysis} 
\label{sec:data_analysis}

\subsection{Galaxy sample} 
\label{sec:data}

We utilise the following works for our analyses: (1) \citet[][hereafter SG16]{Savorgnan2016A}, (2) \citet[][hereafter D+19]{Davis2019}, (3) \citet[][hereafter S+19]{Sahu2019A}, and (4) \citet[][hereafter H+22]{Hon2022}. 
These studies have performed physically-motivated\footnote{Rather than blindly fitting 2, 3, or 4 S\'ersic components, we inspect each image and fit for specific physical components seen in the data.}, multicomponent decompositions of local galaxies of various morphology.
The decomposition in D+19, S+19, and H+22 are performed using \texttt{Profiler} \citep{Ciambur2016}, while SG16 used their own programme "profilterol". 

Among the data from the four works, there are some repeating galaxies.
There is one galaxy in D+19 (NGC~3368) and four galaxies in S+19 (NGC~3665, NGC~4429, NGC~4526, and NGC4649) overlapping with H+22.
We adopt the newer decomposition from H+22 and remove the repeating data points from the previous works.
In total, there are 241 unique galaxies in these data sets.
See the respective papers for the structural parameters of individual galaxies.
The followings are brief descriptions of each.

\subsubsection{Savorgnan \& Graham (2016)} 
\label{sec:Savorgnan_et_al}

SG16 analyzed 66 galaxies with dynamically measured supermassive black hole (SMBH) mass, $M_\mathrm{BH}$, from \citet{Greenhill2003,Graham2013,Rusli2013}. 
It contains 47 ETGs and 19 "early-type" spiral galaxies.
The analysis was performed on $3.6~\mu \rm m$ \textit{Spitzer}/IRAC (InfraRed Array Camera) images \citep[][]{Fazio2004,Werner2004A} (see SG16, their Section~2.1). 
In their subsequent paper, \citet{Savorgnan2016B} convert the $3.6~\mu \rm m$ luminosity into stellar mass using a constant mass-to-light ($M_{*}/L_{3.6}$, or $\Upsilon_{3.6}$) ratio of $0.6$ \citep[][]{Meidt2014}\footnote{\cite{Meidt2014} assumes a \citet{Chabrier2003} initial mass function (IMF), a \citet{Bruzual2003} stellar synthesised stellar population model (SSP), and an exponentially declining stellar formation history (SFH).}.

\subsubsection{Davis et al. (2019)}
\label{sec:Davis_et_al}

D+19 performed multicomponent decompositions for 43\footnote{Three of which are bulgeless galaxies.} late-type galaxies (LTGs).
The sample was selected based on a compilation from \citet{Davis2017}: a set of spiral galaxies with their SMBH mass measured via proper stellar motion, stellar dynamics, gas dynamics, and stimulated astrophysical masers (see their Table~1).
The spiral galaxy data set consists of the following morphological type: 10 SA, 12 SAB, and 22 SB galaxies \citep[defined by RC3][]{de_Vaucouleurs1991rc3}\footnote{Four of which were discarded by D+19 in the decomposition process.}.
D+19 analysed the galaxy structures primarily using $3.6~\mu \rm m$ images. 
D+19 analyzed the \textit{Spitzer} Survey of Stellar Structure in Galaxies \citep[$\rm S^{4}G$,][]{Sheth2010} and the \textit{Spitzer} Heritage Archive (SHA)\footnote{\url{http://sha.ipac.caltech.edu}}.
Alternative data from the $F814W$-band in \textit{Hubble Space Telescope} (HST)\footnote{\url{https://mast.stsci.edu/}} images and $K_\mathrm{s}$-band images from the Two Micron All Sky Survey (2MASS) Large Galaxy Atlas \citep[LGA,][]{Jarrett2003} were used in the absence of $\textit{Spitzer}$ images or when greater special resolution was required (see Section~2.2 in D+19).
In LTGs, because the dust in the disc glows in the infrared, D+19 used a smaller stellar mass-to-light ratio $M_*/L_{3.6} = 0.453$, in accordance with data from \citet{Querejeta2015}, to compensate for the potential overestimation of stellar mass due to non-stellar luminosity in LTGs (see Section~2.8 in D+19)\footnote{D+19 used the median $(L_{*}/L_\mathrm{obs})_\mathrm{IRAC1} = 0.755$ from Fig.~10 of \citet{Querejeta2015} and the color-independent $M_*/L_\mathrm{IRAC1} = 0.60$ from \citet[][]{Meidt2014} to estimate the median value of $M_*/L_{3.6} = 0.453$. See Section~2.8 in D+19 for details.}.
For the sake of consistency, we only use their 26 galaxies with \textit{Spitzer} images.

\subsubsection{Sahu et al. (2019)} 
\label{sec:Sahu_et_al}

Building upon the 47 ETGs from SG16, S+19 added 41 more ETGs, including seven remodelled ETGs (A3565~BCG, NGC~524, NGC~2787, NGC~1374, NGC~4026, NGC~5845, and NGC~7052). 
For these seven remodelled galaxies, we use the parameters provided by S+19.
Among the newly added 41 galaxies, based on their decomposition, are 15 E, 3 ES, and 23 S0 galaxies, respectively. 
S+19 data sources are from: $3.6~\mu \rm m$-band images in \textit{Spitzer}/IRAC \citep{Sheth2010}, r-band images in Sloan Digital Sky Survey \citep[SDSS,][]{York2000}, $K_s$-band images from 2MASS/LGA \citep[][]{Jarrett2003}, and $F814W$-band in SHA.
As was done with the D+19 data, we only use the \textit{Spitzer} galaxies from S+19, which amounts to 33 galaxies.

Similar to what was done in SG16, the spheroid $3.6~\mu \rm m$ luminosity from \textit{Spitzer}/IRAC and SHA images are converted into stellar mass with a constant mass-to-light ratio $M_*/L_{3.6} = 0.6$ \citep[][see their Section~3.3 for details]{Meidt2014}.

\subsubsection{Hon et al. (2022)} 
\label{sec:Hon_et_al}

H+22 selected a mass- and volume-limited sample of massive ($M_\mathrm{*,gal} > 10^{11}~\rm M_{\odot}$) galaxies at $\rm Distance < 110~Mpc $ from the NASA-Sloan catalogue\footnote{\url{http://www.nsatlas.org/}} based on SDSS photometry \citep{York2000,Aihara2011_SDSSDR8} in the $i$-band. 
The H+22 sample contains 103 massive galaxies with $M_*/\rm M_{\odot}>6.7\times10^{10}$ (based on the \citet{Roediger2015} $M_{*}/L_{i}$ ratio) but is otherwise nondiscriminatory to galaxy morphology. 
According to H+22's decomposition, the galaxies are assigned new morphologies.  
There are 13 ellipticals (E+ES)\footnote{Here, the elliptical galaxies are composed of two populations: 11 conventional ellipticals (E) and two ellicular (ES) galaxies.
ES galaxies \citep{Liller1966,Savorgnan2016disc,Graham2019class} are ellipitcals that contain an intermediate-scale disc. 
The name "ellicular" ("elliptical+lenticular") was introduced in \citet{Graham2016}.}, 50 lenticular (S0) and 40 spiral (S) galaxies in this sample. 
H+22 use the galaxies' $g$- and $i$-band photometry from SDSS to estimate the $M_*/L_{i}$ ratio.
In this paper, we will use the $M_*/L_{i}$ ratio provided by the \citet[][]{Roediger2015} prescription, which provides agreement with the \textit{Spitzer}-derived masses from the three other works \citep[][see their Section~3.4]{Sahu2019A}
\footnote{The $M_*/L$ ratio utilised by the four works assume a \citet{Chabrier2003} initial mass function and the stellar population model from \citet{Bruzual2003} \citep[see][]{Meidt2014,Roediger2015}.}.
After removing the non-Spitzer galaxies from the previous three studies and combining them with H+22, we have, in total, 202 galaxies for analysis.

\subsection{Surface brightness profile analysis} 
\label{sec:SBanalysis}

The surface brightness profiles of the host galaxies were previously extracted via the isophotal fitting programme \texttt{ISOFIT} \citep{Ciambur2015}.
Built on the isophotal fitting function \texttt{ellipse} in \texttt{IRAF}, it fits a series of concentric quasi-ellipses onto the galaxy image. 
The flux along each quasi-ellipse gives the surface brightness at a given radius. 
Improving upon \texttt{ellipse}, \texttt{ISOFIT} is able to better handle bars and near-edge-on discs by correctly using high-order Fourier harmonic distortions to ellipses rather than just circles \citep[see][ for details]{Ciambur2015}.

The decomposition is informed by the various radial profiles (surface brightness $\mu$, ellipticity $\epsilon$, Position Angle $PA$, harmonic coefficients $B_{4}$, $B_{6}$) of the galaxy isophotes, describing features in the 2D images.
Relevant studies from the literature were also considered when choosing the appropriate components for each galaxy.
In addition to the more apparent structures, such as a bulge, extended disc, and bar, less prominent but important substructures, such as a nuclear disc \citep[][]{Ferrarese1994,van_den_Bosch1998, Scorza1998,Balcells2007}, secondary bar \citep{Buta1993,Shaw1993,Wozniak1995,Jungwiert1997J,Erwin2003}, anase \citep[][]{2007AJ....134.1863M} and rings \citep[][]{Buta1996ring}, were also included in the decomposition \citep[see Section~3 in][]{Hon2022}. 
The different types of disc, namely the Type I regular disc \citep[][]{1940BHarO.914....9P,de_Vaucouleurs1957}, the down-bending Type II disc \citep[][]{van_der_Kruit1981}, the up-bending Type III disc \citep{2001MNRAS.324.1074D, 2002A&A...392..807P}, and the inclined disc model \citep[][]{van_der_Kruit1981} were also considered. 
Because such potentially biasing factors and components have been accounted for, the bulges/spheroids obtained from the decomposition process are more accurate than simple 2-component Bulge+Disc decomposition.
Rather than relying on a statistical approach, e.g., Akaike information criterion or Bayesian information criterion, additional components were only fit if there was evidence, either pre-existing in the literature or in the SDSS imaging data, for their presence. This included features in the ellipticity ($\epsilon$), position angle ($PA$) or harmonic coefficient (e.g. $B_4$, $B_6$) radial profiles. 
For detailed descriptions of the decomposition process, see Sec. 3.3 and Fig. 6 in H+22.

%\hl{In addition, the multicomponent models were convolved using the PSF measured in the same frame for each galaxy.
%The average PSF of the 3.6 $\rm \mu m$ $Spitzer$ images from SG16, D+19, and S+19 are $R_\mathrm{PSF, Spitzer}\lesssim 2$--$2.2 ~\rm arcsec$.
%In the case where $R_\mathrm{e, Sph} < R_\mathrm{PSF, Spitzer}$, those galaxies were not used in our analysis.
%For the SDSS $i-$band images from H+22, the median PSF size is $\sim 1.0 \rm arcsec$ (see their Table C1--C3 and D1--D3).
%The $R_\mathrm{e, Sph}/R_\mathrm{PSF, SDSS}$ ratio is sufficiently large across all four data sources where the median values are $\langle R_\mathrm{e, Sph}/R_\mathrm{PSF} \rangle\sim 36.90, 9.00, 8.90,$ and $4.67$ for SG+16, D+19, S+19, and H+22, respectively. 
%Therefore, we conclude that the extracted bulges are well-resolved, and the spatial resolution does not bias our measurement of the spheroid properties.}

The multicomponent galaxy models were convolved using a Moffat representation of the PSF measured in the same frame as each galaxy.
For the SDSS $i$-band images from H+22, the median FWHM size of the PSF is $\sim 1.0 \arcsec$, with individual measurements listed in Tables C1--C3 of H+22.  
None of the 103 galaxies had $R_{\rm e,sph}/FWHM < 1.0$.  
As noted in section 2.2 of D+19, when $R_{\rm e,Sph}$ was comparable to or smaller than the `seeing' of the \textit{Spitzer} Space Telescope ($\sim 2 \arcsec$), then HST data was used for those galaxies with directly measured black hole masses, which are not included here due to our effort to ensure the use of consistent $M/L_*$ ratios between the samples.  
This was also the case with the Spitzer sample from SG16 and S+19.
That is, the black hole sample used here has \textit{a priori} excluded galaxies with $R_{\rm e,Sph}/FWHM \lesssim 1$, and simulations have repeatedly shown that the S\'ersic model's $R_{\rm e,sph}$ can be recovered under such conditions \citep{Gadotti2008}.

%In particular, because most spiral galaxies from H+22 that contain small low-mass bulges are within $\sim 30\rm~Mpc$, it equates to a PSF size of $\sim0.15~\rm kpc$, which is still sufficiently smaller than these low-mass spheroids (with $R_\mathrm{e, Sph}\sim 0.2$--$0.3~\rm kpc$). 

%Compared to the typical size of the spheroids ($R_\mathrm{e, Sph}\sim 0.1$--$10~\rm kpc$), the PSF is insufficient to bias the estimation of the spheroids' sizes. 

Finally, it is perhaps useful to remark on the nature of the spheroids, whether they are the classical bulges or pseudo-bulges built via secular influence from the disc. 
SG16 stated that they did not attempt to distinguish between classical and pseudo-bulges. 
However, with most of their data being barless S0 galaxies and giant elliptical galaxies, pseudo-bulges are not likely to be present in SG16.
Because D+19 focuses on LTGs, they reported on claims in the literature for 35 pseudo-bulges in their sample of 43.
However, having their galaxies' bar components and (peanut shell)-shaped structures captured and modelled, bar-induced pseudo-bulge structures were effectively subsumed into the bar model.
D+19 note that all their bulges co-define the same $M_\mathrm{BH}$-$M_\mathrm{*, Sph}$ relation.
A tight $M_\mathrm{BH}$-$M_\mathrm{*, Sph}$ relation implies the bulges they extracted are the structures that coevolve with the SMBH, namely the classical spheroids instead of the secularly-built pseudo-bulges.
S+19 modelled the boxy/X/(peanut shell)-shaped structure, as well as the oval (barlens) structure, separate from the (classical) bulges. They used either a Ferrers \citep[][]{Ferrers1877,Sellwood1993Bar} or S\'ersic function \citep[][]{Sersic1968} for these "pseudo-bulge" structures (see in their plots, NGC~4792, NGC~2787, and NGC~4371); that is, the disc-induced, secularly-built pseudo-bulges are captured in either the bar/barlens component.
The same treatment can be seen in H+22.
Therefore, the spheroids described in this work are thought to be classical bulges and not inner discs, which are modelled as such, nor features thrown up by unstable bars.

\subsection{Spheroid model} 
\label{sec:sph_mod}

The spheroids are modelled by either one of two functions: the S\'{e}rsic or the core-S\'{e}rsic function \citep[][]{Graham2003coreSersic}.
The S\'{e}rsic function describes the bulges' intensity profile with three parameters ($n$, $R_\mathrm{e}$, and $I_\mathrm{e}$), such that

\begin{equation} 
\label{eq:Sersic}
I(R)=I_{\rm e} \exp \left\{-b_{n}\left[\left(\frac{R}{R_{\rm e}}\right)^{\frac{1}{n}}-1\right]\right\}, 
\end{equation}
where $n$ is the S\'{e}rsic index or `shape parameter' that depicts the shape of the profile, $R_\mathrm{e}$ is the "effective" half-light radius that encapsulate half the total luminosity, and $I_\mathrm{e}$ is the intensity at $R_\mathrm{e}$.
It is easy to see that $I_\mathrm{0} \equiv I(R=0)= I_\mathrm{e} ~e^{b_\mathrm{n}}$.
The quantity $b_\mathrm{n}$ is a function of $n$, determined by solving the equation $\Gamma(2n) = 0.5\gamma(2n,b_\mathrm{n})$, where $\Gamma$ and $\gamma$ are the complete and incomplete gamma functions, respectively \citep[see][]{Graham2005PASA}. 
For $0.5 < n < 10$, $b_\mathrm{n}$ can be approximated by the following equation: $b_\mathrm{n} \approx 1.9992n-0.3271$ \citep[][]{Capaccioli1989}. 
The total luminosity ($L_\mathrm{tot}$) can be calculated by integrating the intensity over the projected surface of the galaxy. 
We have:
\begin{equation} 
\label{eq:L_tot}
L_{\text {tot}}=\frac{I_{0} R_{\mathrm{e}}^{2} 2 n \pi}{\left(b_{n}\right)^{2 n}} \Gamma(2 n),
\end{equation}
and the surface brightness ($\mu$) at a given radius ($R$) is:
\begin{equation} 
\label{eq:mu_equ}
\mu(R)=-2.5 \log(I(R)).
\end{equation}

The S\'{e}rsic function has a particular noteworthy construct.
The value of $b_n$ is arbitrarily defined in \citet{Sersic1968} to be the value where the parameter, $R_\mathrm{e}$, encompasses half of the total luminosity ($0.5L_\mathrm{tot}$).
When switching to a different scale radius, $R_\mathrm{z}$, encompasses a different fraction $z$ of the total light, and this is achieved by changing $b_n$ to a new constant $b_{n,z}$.
We can convert $R_\mathrm{e}$ to $R_\mathrm{z}$ using equation 22 in \citet{{Graham2019PASA}}:

\begin{equation} 
\label{eq:Rz}
R_\mathrm{z} = \left( \frac{b_{n,z}}{b_\mathrm{n}} \right)^{n} R_\mathrm{e}.
\end{equation}
where $b_{n,\mathrm{z}}$ is a function of $n$ obtained by solving the equation $\Gamma(2n) = z\gamma(2n,b_\mathrm{n})$, and $z$ is the enclosed fraction of light ranging from 0 and 1. 

The scale radius, $R_z$, is often overlooked in the literature.
As one chooses to characterise the spheroid structures with a different $R_z$ instead of $R_\mathrm{e}$, the distribution of $\mu_z$ and projected mass density $\Sigma_z$ will change as well.
As a result, the corresponding scaling relations differ drastically from when $R_\mathrm{e}$ was used.
One might incorrectly assign physical significance to a particular relation without realising that some quantities are arbitrarily defined and unrelated to the underlying physics of galaxy evolution.
We shall reveal how the correlation strength changes for different scaling relations as a function of light fraction $z$ later in Section~\ref{sec:scale_radius}.

Similarly, the corresponding surface brightness at $R_z$, $\mu_\mathrm{z}$, will be:

\begin{equation} 
\label{eq:muz}
\mu_\mathrm{z} = \mu_{0}+2.5\frac{b_{n,z}}{\mathrm{ln}(10)},
\end{equation}
where $\mu_{0}$ is the central surface brightness.
The average surface brightness, $\langle\mu\rangle_{z}$, within $R_z$ is:

\begin{equation} 
\label{eq:average_muz}
\langle\mu\rangle_{z}=\mu_{z}-2.5 \log \bigg(\frac{ 2 z n \mathrm{e}^{b_{n, z}}}{\left(b_{n, z}\right)^{2 n}} \Gamma(2 n)  \bigg) = \mu_{z} -2.5\log(B_z(n)),
\end{equation}
where, here, we name the function $B_z(n)$ the `shape function'\footnote{ \citet{Graham2019PASA} refer to this function as $f(n)$ (see their equation~20).} of a S\'ersic profile:
\begin{equation} 
\label{eq:average_Bnz}
B_z(n) \equiv \frac{ 2 z n \mathrm{e}^{b_{n, z}}}{\left(b_{n, z}\right)^{2 n}} \Gamma(2 n). 
\end{equation}
Note that $B_z(n)$ is required to calculate the total luminosity of a S\'ersic profile with $L_\mathrm{tot} = I_{0}R_{z}^{2}B_z(n)e^{-b_\mathrm{n,z}}\pi$. 

The surface brightness, $\mu_z$, can be converted into a physical quantity, the projected mass density, $\Sigma_\mathrm{z}$:
\begin{equation} 
    \begin{split}\label{eq:sigma0_law}
-2.5 \log \left(\Sigma_{\mathrm{z}}\left[\mathrm{M}_{\odot} \mathrm{pc}^{-2}\right]\right)= & \mu_{\mathrm{z}}-\mathrm{DM}-\mathfrak{M}_{\odot, \lambda}-2.5 \log \left(1 / \mathrm{s}^{2}\right)\\
&-2.5 \log \left(\Upsilon_{\lambda}\right),
    \end{split}
\end{equation}
where $\mu_z$ is the surface brightness at $R_z$; $\rm DM$ is the distance modulus; $\mathfrak{M}_{\odot,\lambda}$ is the absolute magnitude of the Sun at wavelength $\lambda$; $s$ is the physical size scale in $\rm pc~arcsec^{-1}$; and  $\Upsilon_{\lambda}$ is the mass-to-light ratio at wavelength $\lambda$ \citep[][see their equation A5 for details]{Sahu2022}.

Some massive galaxies/spheroids have surface brightness profiles which deviate from the S\'{e}rsic model because of their core-deficit \citep{King1978, Byun1996, Trujillo2004}.
The merging of binary SMBHs ejects the neighbouring stars outwards through gravitational slingshots \citep[][]{Saslaw1974, Begelman1980}.
As a result, the surface brightness profile of the core is shallow out to some "break-radius", $R_\mathrm{b}$, and conforms to a regular S\'ersic model in the outer region. 
For this special class of spheroids, they are modelled with the six parameters ($n, R_\mathrm{e}, R_\mathrm{b}, I_\mathrm{b},\alpha, \rm and~\gamma$) "core-S\'{e}rsic" function\footnote{The core-S\'{e}rsic function is an empirical model combining a power-law to describe the depleted core with the S\'{e}rsic function.} \citep{Graham2003coreSersic}:

\begin{equation} 
\label{eq:core-Sersic}
I(R)=I^{\prime}\left[1+\left(\frac{R_\mathrm{b}}{R}\right)^{\alpha}\right]^{\frac{\gamma}{\alpha}} \exp \left\{-b_{n}\left[\frac{R^{\alpha}+R_\mathrm{b}^{\alpha}}{R_\mathrm{e}^{\alpha}}\right]^{\frac{1}{\alpha n}}\right\},
\end{equation}
where $R_\mathrm{b}$ is the "break-radius", $\gamma$ describes the slope of the inner power-law profile, and $\alpha$ describes the sharpness of the transition between the inner (power-law) and outer (S\'ersic) regimes. 
The intensity $I'$ can be expressed as such:

\begin{equation} 
\label{eq:Iprime}
I^{\prime}=I_\mathrm{b} 2^{-(\gamma / \alpha)} \exp \left[b_\mathrm{n}\left(2^{1 / \alpha} R_\mathrm{b} / R_\mathrm{e}\right)^{1 / n}\right], 
\end{equation}
where $I_\mathrm{b}$ is the intensity at the break-radius.
See Fig. 9 and 10 in \citet[][]{Graham2003coreSersic} for the differences in S\'ersic and core-S\'ersic functions.

In what follows, we present the spheroid scaling relations, as well as their structural parameters as described by these two models.
Unless otherwise stated, all parameters (e.g., $n$, $\mu_{\rm e}$,) shown in this work are calculated using radii, $R_\mathrm{eq}$, equivalent to the geometric-mean of the major ($a$) and minor ($b$) axis: $R_\mathrm{eq} = \sqrt{ab}$.
This implicitly accounts for the ellipticity of the isophotes.

\section{Results} 
\label{sec:Result}

This section is structured as follows:
\begin{enumerate}
\item We first establish the correlation strength of various relations involving the spheroid structural parameters, $n_\mathrm{Sph},~\mu_\mathrm{0,Sph}$, and $R_\mathrm{e,Sph}$, absolute magnitude, $\mathfrak{M}_\mathrm{Sph}$ in Section~\ref{sec:Struc_para} and \ref{sec:Corr_among_para} using H+22 sample.

\item Afterward, we demonstrate how these distributions change while using alternate scale radii, $R_z$ (see equation~\ref{eq:Rz}), in Section~\ref{sec:scale_radius}.
We also explore how the effective surface brightness (equation~\ref{eq:muz}) changes with $\mathfrak{M}_\mathrm{Sph}$ for a range of fractions, $z$.

\item In Section~\ref{sec:mag_to_mass}, we convert the spheroid surface brightness and  magnitude into physical quantities: $\Sigma_\mathrm{z, \rm Sph}$, $\langle\Sigma\rangle_{z, \rm Sph}$, and  $M_\mathrm{*, Sph}$, where $\Sigma_{z, \rm Sph}$ and $\langle \Sigma \rangle_{z,\rm Sph}$ are the stellar surface mass density at and within $R_{z, \rm Sph}$, and $M_\mathrm{*,Sph}$ is the spheroid stellar mass.
This enables us to fold in the sample from SG16, D+19, and S+19 in to the analysis.
In Section~\ref{sec:corr_trend}, we explore the correlation strength of various quantities: $n_\mathrm{Sph}$, $R_{z,\rm Sph}$, $\Sigma_{z,\rm Sph}$, $\langle\Sigma\rangle_{z, \rm Sph}$, and $B_{z, \rm Sph}(n_\mathrm{Sph})$ with the spheroid stellar mass, $M_\mathrm{*,Sph}$, under the assumption of different scale radii.

\item We then fit the spheroid size--mass ($R_\mathrm{e,Sph}$--$M_\mathrm{*,Sph}$) relation in Section~\ref{sec:size_mass}.
Due to its strong linearity, the size--mass relation is arguably the most important scaling relation.
Section~\ref{sec:additional_relations} presents a few additional fitted scaling relations: $n_\mathrm{Sph}$--$M_\mathrm{*,Sph}$, $\Sigma_\mathrm{0,Sph}$--$n_\mathrm{Sph}$, and $n_\mathrm{Sph}$--$R_\mathrm{e,Sph}$ that are supplementary to the size--mass relation.

\end{enumerate}

\subsection{Absolute Magnitude versus S\'ersic index, central surface brightness, and effective radius} 
\label{sec:Struc_para}

\begin{figure*}
\centering
	\includegraphics[clip=true, trim=2mm 2mm 4mm 0mm, width=1.0\textwidth]{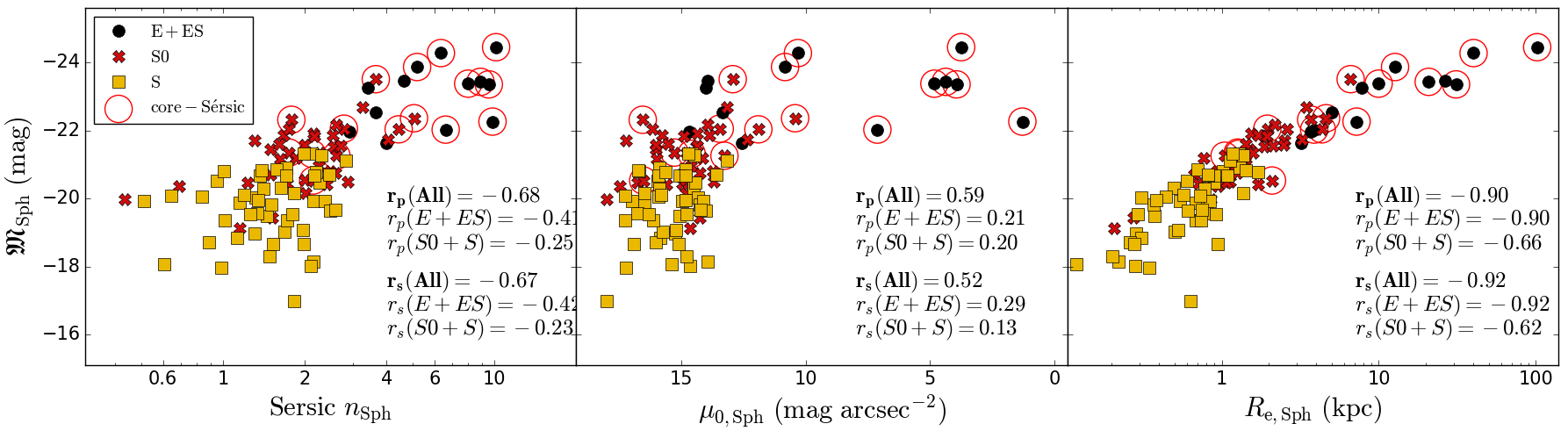}
    \caption{Spheroid absolute magnitude, $\mathfrak{M}_\mathrm{Sph}$ ($i$-band, AB), versus the structural parameters ($n_\mathrm{Sph}$, $\mu_\mathrm{0,Sph}$, and $R_\mathrm{e,Sph}$, from the equivalent axis) from the decomposition in H+22. 
    The $\mu_0$ value of the core-S\'{e}rsic spheroids are derived from extrapolations towards the centre, based on the three parameters ($R_\mathrm{e}$, $\mu_\mathrm{e}$, and $n$) in the S\'ersic component of the core-S\'ersic model. Essentially, they depict $\mu_0$ as if there is no core-deficit. 
    Spheroids modelled by the core-S\'{e}rsic model are marked with red open circles ($\bigcirc$).
    The Pearson correlation coefficients ($r_\mathrm{p}$) and Spearman rank-order correlation coefficient ($r_\mathrm{s}$) for each relation are displayed. 
    }
    \label{fig:M_n_mu}
\end{figure*}

Fig.~\ref{fig:M_n_mu} shows the distribution of the total spheroidal absolute magnitude, $\mathfrak{M}_\mathrm{Sph}$, versus the S\'{e}rsic index, $\log(n_\mathrm{Sph})$, central surface brightness, $\mu_\mathrm{0,Sph}$, and effective radius, $\log(R_\mathrm{e,Sph})$, from the sample in H+22.
The core-S\'{e}rsic spheroids, highlighted by red circles, are universally bright with $\mathfrak{M}_\mathrm{Sph} < -21~\rm mag$ in the $i$-band (AB mag).
The distribution of spheroid parameters separates E+ES galaxies and spheroids from S0 and S galaxies.
In the $\mathfrak{M}_\mathrm{Sph}-\mathrm{log}(n_\mathrm{Sph})$ plot (left-hand panel), most spheroids residing at $n_\mathrm{Sph} > 3.5$ and $ \mathfrak{M}_\mathrm{Sph} < -21~\rm mag$ are elliptical (E+ES) galaxies. 
Bulges coming from S0 and S galaxies concentrate in $ -21 < \mathfrak{M}_\mathrm{Sph}/\mathrm{mag} <- 18$ and $0.4< n_\mathrm{Sph} < 3.5$.
In the middle panel of Fig.~\ref{fig:M_n_mu}, the $\mathfrak{M}_\mathrm{Sph}$--$\mu_\mathrm{0,Sph}$ plot presents a rather different distribution.
S0 and S spheroids are concentrated between $ 12 \lesssim \mu_\mathrm{0,Sph}/\rm mag~arcsec^{-2}  \lesssim 20$ and $-23 < \mathfrak{M}_\mathrm{Sph}/\rm~ mag  < -18 $, while the E+ES galaxies span a different range, at  $ 15 \lesssim \mu_\mathrm{0,Sph}/\rm mag~arcsec^{-2}  \lesssim 0$ and $-23 < \mathfrak{M}_\mathrm{Sph}  < -17 \rm~mag$.
These differences likely reflect a major merger origin for the ES/E galaxies, inflating their $R_\mathrm{e}$ and luminosity as the disc mass of the progenitors is turned into spheroid mass and resulting in larger S\'ersic indices and brighter (extrapolated) $\mu_0$ values.
In practice, the damage caused by coalescing binary SMBHs erodes the central stellar phase space and reduces the observed $\mu_0$ value.
The disc stars from the progenitor galaxies contribute to the long tail of their high-$n$ profile.
The $\mathfrak{M}_\mathrm{Sph}$--$\log(R_\mathrm{e, Sph})$ relation, shown in the right-hand panel, exhibits a strong log-linear trend, with good agreement across the E+ES, S0 and S bulges. 
At $R_\mathrm{e,Sph}\gtrsim 3$--$4 \rm~ kpc$, the spheroid population is dominated by E+ES galaxies. 

Studies have reported linear scaling relations for ETGs in the  $\mathfrak{M}_\mathrm{gal}$--$\log(n_\mathrm{gal})$ and $\mathfrak{M}_\mathrm{gal}$--$\mu_\mathrm{0,gal}$ planes \citep[][]{Caon1993,Young1994,Graham1996,Graham2001A,Ferrarese2006a,Savorgnan2013,Janz2014}.
For instance, using a single S\'ersic function to model ETGs in the $B$-band \citep[][]{Binggeli1998,Caon1993,D'Onofrio1994, Stiavelli2001,Graham2003}. 
Researchers have investigated the empirical  $\mathfrak{M}_\mathrm{gal}$--$\log(n)$ and $\mathfrak{M}_\mathrm{gal}$--$\mu_\mathrm{0,gal}$ for massive ETGs.
In \citet[][see their Fig.~6]{Graham2013}, both distributions exhibit a clear linear relation (see also their equation 6 and 7), and they use these empirical relations to predict the ETG size--luminosity relation in the $B$-band. 

However, for the \textit{spheroid} structural parameters shown in Fig.~\ref{fig:M_n_mu}, the $\mathfrak{M}_\mathrm{Sph}$--$\mu_\mathrm{0,Sph}$ and $\mathfrak{M}_\mathrm{Sph}$--$\log(n_\mathrm{Sph})$ relations are not obviously log-linear.
The $\mathfrak{M}_\mathrm{Sph}$--$\log(n_\mathrm{Sph})$ relation appears to have a wide range of scatter, and the $\mathfrak{M}_\mathrm{Sph}$--$\mu_\mathrm{0, Sph}$ distribution has a prominent "bend" at $\mathfrak{M}_\mathrm{Sph}\sim-22~\rm mag$ (AB mag), with E+ES galaxies having much brighter central surface brightness compared to the S0 and S bulges.
Interestingly, our $\mathfrak{M}_\mathrm{Sph}$--$\mu_\mathrm{0,Sph}$ distribution is similar to \citet[][]{Watkins2022}'s $\mathfrak{M}_\mathrm{gal}$--$\mu_\mathrm{0,gal}$ relation for their ETGs (see their Fig.~14).
While their distribution is also clearly not linear, by excluding galaxies with high S\'ersic index ($n\gtrsim7$), they provide a linear  $\mathfrak{M}_\mathrm{gal}$--$\mu_\mathrm{0, gal}$ relation (see their equations 15--16).
Here, we first test the linearity among the three distributions shown in our Fig.~\ref{fig:M_n_mu}.

We computed the Pearson correlation coefficient ($r_\mathrm{p}$) and the Spearman rank-order correlation coefficient ($r_\mathrm{s}$) for each distributions (listed in Fig.~\ref{fig:M_n_mu}).
The coefficient $r_\mathrm{p}$ depicts how well a linear relation describes our data and $r_\mathrm{s}$ describes how well the relation approximates a monotonic increasing trend.
For our spheroids, both $r_\mathrm{p}$ and $r_\mathrm{s}$ are similar across all parameters.
The $\mathfrak{M}_\mathrm{Sph}$--$\log(R_\mathrm{e,Sph})$ relation presents the highest positive correlation among the three, with $r_\mathrm{p}=-0.90$ and $r_\mathrm{s}=-0.92$.
Note that the coefficient is negative because a more luminous object has a lower absolute magnitude.
Following the observation above on the different locations of elliptical and disc galaxies in these scaling diagrams, we subdivided our sample into E+ES and S0+S galaxies to check if there is any deviation between the two groups.
The Pearson and Spearman coefficients for ellipticals, $r_\mathrm{p}(\rm E+ES)$ and $r_\mathrm{s}(\rm E+ES)$, and disc-galaxies, $r_\mathrm{p}(\rm S0+S)$ and $r_\mathrm{s}(\rm S0+S)$, are shown in Fig.~\ref{fig:M_n_mu}.
Note that, in particular, in the $\mathfrak{M}_\mathrm{Sph}$--$\mu_\mathrm{0,Sph}$ plane, disc-galaxies exhibit a different pattern than the ellipticals, with $r_\mathrm{p}(\rm S0+S)=0.20$ and $r_\mathrm{p}(\rm E+ES)=0.21$.

\subsection{Correlation between \texorpdfstring{$n_\mathrm{Sph},  \mu_\mathrm{0,Sph},$ and $R_\mathrm{e,Sph}$}{}}
\label{sec:Corr_among_para}

\begin{figure*}
\centering
	\includegraphics[clip=true, trim=2mm 2mm 2mm 0mm, width=0.95\textwidth]{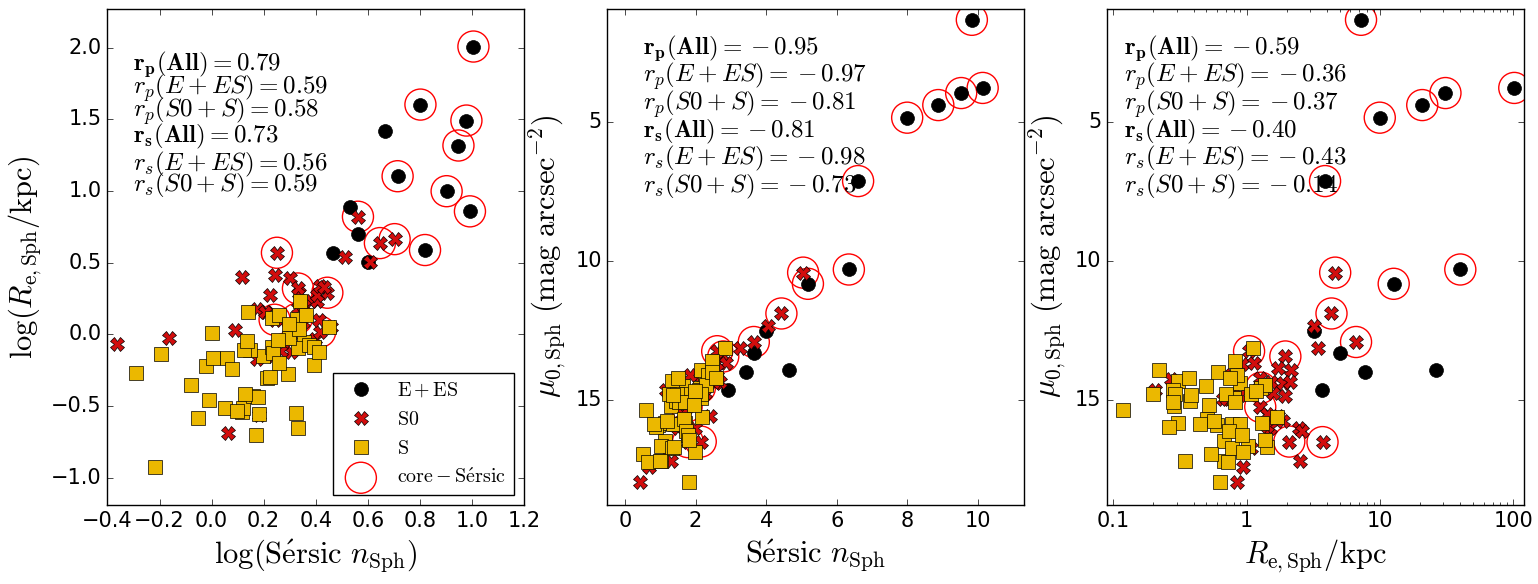}
    \caption{Left-hand panel: The H+22's spheroid effective radius, $\log(R_\mathrm{e,Sph})$, versus S\'ersic index, $\log(n_\mathrm{Sph})$.
    Middle panel: central surface brightness, $\mu_\mathrm{0,Sph}$ ($i$-band, AB), versus S\'ersic index, $n_\mathrm{Sph}$.
    Right-hand panel: the central surface brightness, $\mu_\mathrm{0,Sph}$, versus effective radius, $\log(R_\mathrm{e,Sph})$.}
    \label{fig:L_plot}
\end{figure*}

Beyond comparing structural parameters to the spheroid magnitude, we also study how these parameters relate to each other.
Fig.~\ref{fig:L_plot} shows the distribution of H+22's spheroids in the $\log(R_\mathrm{e,Sph})$--$\log(n_\mathrm{Sph}$), $\mu_\mathrm{0,Sph}$--$n_\mathrm{Sph}$ and $\mu_\mathrm{0,Sph}$--$\log(R_\mathrm{e,Sph})$ planes.

Among the three relations, the  distribution in the $\mu_\mathrm{0,Sph}$--$n_\mathrm{Sph}$ plane shows the strongest correlation ($r_p(All)=-0.95$ and $r_s(All)=-0.81$).
It is remarkably linear, especially for the E+ES galaxies ($r_p(E+ES)=-0.97$) due to extrapolation of the S\'ersic model, with high-$n$ profiles giving bright $\mu_0$ values.
In the $\log(R_\mathrm{e,Sph})$--$\log(n_\mathrm{Sph}$) plane, the spheroids exhibit a moderately high correlation ($r_p(All), r_s(All) \approx 0.79, 0.73$).
The shape of the distribution is somewhat reminiscent of that seen in the $\mathfrak{M}_\mathrm{Sph}$--$\log(n_\mathrm{Sph})$ diagram in Fig.~\ref{fig:M_n_mu}.
Given that $\mathfrak{M}_\mathrm{Sph}$ and $\log(R_\mathrm{e,Sph})$ are strongly correlated, forming essentially a log-linear relation, it is not surprising that the $\log(R_\mathrm{e,Sph})$--$\log(n_\mathrm{Sph})$ distribution resembles the $\mathfrak{M}_\mathrm{Sph}$--$\log(n_\mathrm{Sph})$ distribution.
The $\mu_\mathrm{0,Sph}$--$\log(R_\mathrm{e,Sph})$ distribution does not exhibit any obvious scaling relation beyond a broad trend.
E+ES galaxies and bulges from S0+S galaxies occupy different regions of the plot with most S0+S bulges within $ 0.1 < R_\mathrm{e,Sph}/\rm kpc < 3$ and $13 <\mu_\mathrm{0,Sph}/\rm mag~arcsec^{-2} < 18.5$.
A spheroid effective radius, $R_\mathrm{e,Sph}$, is not particularly useful in predicting the central surface brightness, $\mu_\mathrm{0,Sph}$.

Two sets of spheroid parameters: $\mathfrak{M}_\mathrm{Sph}$--$\log(R_\mathrm{e, Sph})$ and $\mu_\mathrm{0, Sph}$--$n_\mathrm{Sph}$ have the highest correlation among any pairs and, therefore, can be used as the primary scaling relations to predict spheroid properties reliably. 
In contrast, $\log(R_\mathrm{e,Sph})$--$\log(n_\mathrm{Sph})$ and $\mathfrak{M}_\mathrm{Sph}$--$n_\mathrm{Sph}$ relations are moderately correlated with a varying range of scatter.
As such, $R_\mathrm{e,Sph}$ and $\mathfrak{M}_\mathrm{Sph}$ can be used as a supplementary predictor for the spheroid shape, $n_\mathrm{Sph}$ if $\mu_\mathrm{0,Sph}$ is not available. 

The $\mu_\mathrm{0, Sph}$--$n_\mathrm{Sph}$ relation is one of the more prominent scaling relations among spheroid parameters.
In \citet{Khosroshahi2000B}, using both the bulges from two component decompositions of disc galaxies and E galaxies, they found that the $\mu_\mathrm{0,Sph}$--$\log(n_\mathrm{Sph})$ relation, has a high correlation factor of $r_p\sim -0.88$.
It suggests the shape of the spheroid's light profile, measured by the S\'ersic index, $n_\mathrm{Sph}$, dictates the (at least extrapolated) value of its central surface brightness, $\mu_\mathrm{0, Sph}$, or vice versa.
However, unlike \citet{Khosroshahi2000B}, we found our spheroid $\mu_\mathrm{0,Sph}$--$n_\mathrm{Sph}$ relation to be linear instead of the $\mu_\mathrm{0,Sph}$--$\log(n_\mathrm{Sph})$ relation ($r_p(All) = -0.95$, see Fig.~\ref{fig:L_plot}).
In the $\mu_\mathrm{0,Sph}$--$\log(n_\mathrm{Sph})$ plane, our spheroids have a \textit{curved} relation.

\subsection{Alternate scale radii \texorpdfstring{$R_z$}{} and the associated scale intensities}
\label{sec:scale_radius}

From this Section onward, we examine the spheroid parameters using alternate scale radii $R_{z, \rm Sph}$.
The local ETGs' size-mass relation changes slopes at a bending point near the $B$-band magnitude, $\mathfrak{M}_{B} \sim -18 \rm~mag$, when using $R_\mathrm{e}$ \citep[e.g.,][]{2013pss6.book...91G}.
Such a feature had been used to argue in favour of distinct formation scenarios between two classes of objects \citep[][]{Kormendy2009ApJS,Tolstoy2009,Kormendy2012ApJS,Somerville2015,Kormendy2016}.
However, an extensive review by \citet{Graham2019PASA} pointed out that the curvature in the galaxy size-mass relation is artificial.
As shown in their Section~7 and Fig.~3, the shape of the curved size--mass relation varies drastically when radii that encompass a different, say, 10 or 90 per cent of the light, are chosen, i.e. different scale radii.
Moreover, the magnitude at the bend point that supposedly divides ETGs and dETGs galaxies also change accordingly, and therefore the alleged division at $\mathfrak{M}_{B} = -18~\rm mag$ carries no physical meaning.
To avoid misinterpretation, we shall bare in mind the effect of chosen scale radii on the distribution of the spheroid parameters.

%For simplicity, we sometimes drop the subscripts `Sph'. 
%However, all parameters pertain to the spheroid component of the galaxies

\subsubsection{ \texorpdfstring{$R_\mathrm{0.05}~\&~R_\mathrm{0.95}$}{} and \texorpdfstring{$\mu_\mathrm{0.5}~\& ~\mu_\mathrm{0.95}$}{}}
\label{sec:R_variation}

\begin{figure*} 
\centering
	\includegraphics[ width=0.8\textwidth]{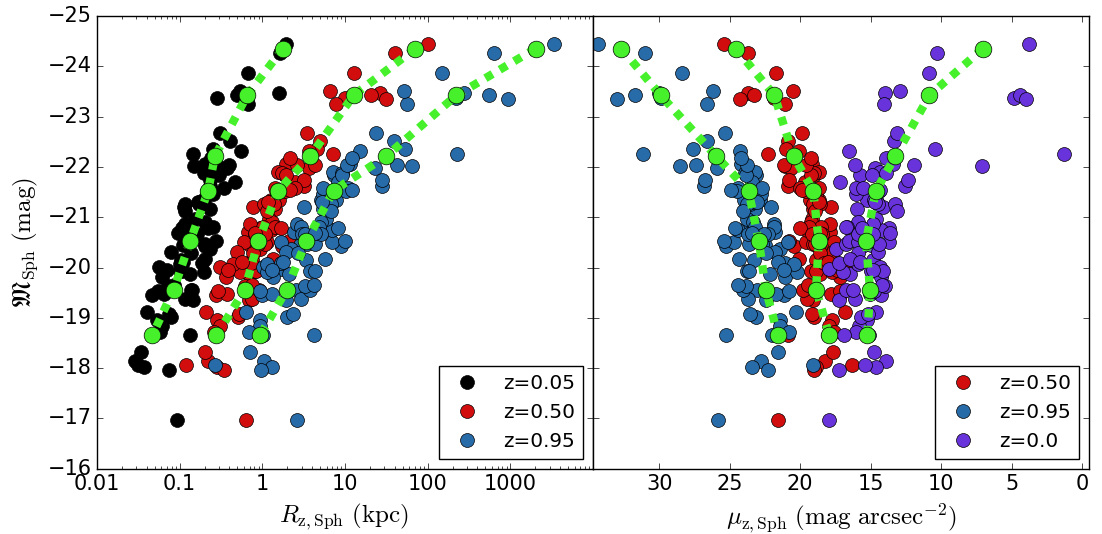}
    \caption{
    Left-hand panel: Spheroids' $i$-band absolute magnitude, $\mathfrak{M}_\mathrm{Sph}$ (in AB mag) versus the equivalent-axis radius, $R_{z,\rm Sph}$, and 
    Right-hand panel: Surface brightness, $\mu_{z,\rm Sph}$, relation using H+22's data. 
    The plot presents the different radius, $R_{z, \rm Sph}$, and surface brightness, $\mu_{z, \rm Sph}$, defined by the different fraction (z, from 0 to 1) of light encapsulated within the radius $R_{z, \rm Sph}$ (see Equation~\ref{eq:Rz} and \ref{eq:muz}). 
    The radius $R_{0.05, \rm Sph}$, $R_{0.5, \rm Sph}$ ($= R_\mathrm{e, Sph}$), and $R_{0.95, \rm Sph}$, correspond to the radius capturing 5 percent, 50 percent, and 95 percent of the spheroid light, are depicted by black, red, and blue points, respectively. 
    Similarly, the surface brightness $\mu_{0, \rm Sph}$, $\mu_{0.5, \rm Sph}$ ($=\mu_\mathrm{e, Sph}$), and $\mu_{0.95, \rm Sph}$ are shown as well. 
    Note that $\mu_{0, \rm Sph}$ is shown in purple points. 
    The bright green dashed lines are drawn using the median value of $R_{z,\rm Sph}$ and $\mu_{z,\rm Sph}$ within the bins defined by a $\Delta\mathfrak{M}_\mathrm{Sph} = 1 $ mag width, from $\mathfrak{M}_\mathrm{Sph}=-18$ to $-25$ ($i$-band AB mag).}
    \label{fig:diff_R}
\end{figure*}

In this section, we develop the spheroid $\mathfrak{M}_\mathrm{Sph}$--$\mu_\mathrm{Sph}$ and $\mathfrak{M}_\mathrm{Sph}$--$R_\mathrm{Sph}$ relations, with a different definition for the radial scale parameter, $R_{z, \rm Sph}$, and thus also the intensity scale parameter ($I_z:\mu_z \equiv -2.5 \log I_z$).
The canonical approach with the S\'ersic model is to only use the effective half-light radius $R_\mathrm{e}$.
We examine how the various relations change when a different fraction ($z$, from 0 to 1) of the total luminosity, $L_\mathrm{tot}$, is used to define the equally-valid alternate scale radii, $R_\mathrm{z}$.

In the left-hand panel of Fig.~\ref{fig:diff_R}, we show $\mathfrak{M}_\mathrm{Sph}$--$\log(R_{z,\rm Sph})$ using a radius $R_{z, \rm Sph}$ encapsulating 50 percent ($z=0.5$, red points), 5 percent ($z=0.05$, black points) and 95 percent ($z=0.95$, blue points) of the light. 
We illustrate the trends for each relation with green dashed lines.
The green points are the median value within an absolute magnitude interval of $\Delta\mathfrak{M}_\mathrm{Sph}=1~\rm mag$ from $\mathfrak{M}_\mathrm{Sph} = -18\rm~mag$ to $-25 \rm~mag$.
In the right-hand panel of Fig.~\ref{fig:diff_R}, the $\mathfrak{M}_\mathrm{Sph}$--$\mu_{z,\rm Sph}$ distributions are presented, where $\mu_{z,\rm Sph}$ is the surface brightness at  $R_{z,\rm Sph}$.
For $z=0$ (purple points), it recreates the same plot as shown in the middle panel of Fig.~\ref{fig:M_n_mu}.  
When $z=0$, the $\mathfrak{M}_\mathrm{Sph}$--$\mu_{z, \rm Sph}$ relation presents a positive trend, i.e., a brighter spheroid has a more luminous central surface brightness value. 
For $z=0.5$ and $z=0.95$, both relations change directions and have their absolute magnitude negatively correlating with the surface brightness, $\mu_\mathrm{0.5,Sph}$ (or $\mu_\mathrm{e}$) and $\mu_\mathrm{0.95,Sph}$.
In both panels, the spheroids have a continuous relations between
$\mathfrak{M}_\mathrm{Sph}\sim-18$ to $-25$ (in $i$-band AB mag). 
The spheroid population appears to be continuous regardless of the radius $R_{z,\rm Sph}$ used.

\subsection{From magnitude to mass}
\label{sec:mag_to_mass}

\subsubsection{\texorpdfstring{$M_{\rm *,Sph}$--$n_\mathrm{Sph}$}{}, \texorpdfstring{$M_{\rm *, Sph}$--$\Sigma_{z,\rm Sph}$}{} and \texorpdfstring{$M_{\rm *, Sph}$--$R_{z,\rm Sph}$  relations}{}}
\label{sec:sigma0}

In an effort to extract physical meaning from these relations, as well as comparing the data from H+22 with SG16, D+19 and S+19, we converted the spheroids' absolute magnitude, $\mathfrak{M}_\mathrm{Sph}$, into stellar mass, $M_\mathrm{*,Sph}$, and central surface brightness, $\mu_0$, into central projected mass density, $\Sigma_\mathrm{0,Sph}$.
Using the data from the four sources, we plotted the spheroid stellar mass, $\log(M_\mathrm{*,Sph})$, versus S\'ersic index, $\log(n_\mathrm{Sph})$,  $\log(\Sigma_\mathrm{0,Sph})$, and $\log(R_\mathrm{e,Sph})$ in the left-hand, middle, and right-hand panel in Fig.~\ref{fig:mass_sigma0}, respectively.

\citet[][their Fig.22]{Kelvin2012} report how the {\em galaxy} half-light radii change with wavelength, which is related to the bulge/disc transition, such that the (small) different colours between the bulge and disc can produce a change in the {\em galaxy} size as a function of wavelength \citep[][]{Kennedy2016}.
\citet[][]{Kelvin2012} report that the drop in the median {\em galaxy} size from the $i$-band (0.6 micron) to the $J$, $H$, and $K$-band (1-2 micron) is $\sim$4.5 kpc to 3.5 kpc, before the decline in size stabilises beyond 1 micron.   
This is a 22\% drop, or 0.11 dex, for the {\em galaxy} half-light size. 
\citet{Vulcani2014} performed the same analysis, and for their `red galaxies', the median {\em galaxy} size dropped from 5.5 kpc to 4 kpc (0.14 dex) when going from the $i$-band (0.6 micron) to the $H$-band (1.6 micron).  
However, what is required is the change in the bulge/spheroid size rather than the change in the galaxy (bulge+disc+bar) size.
As discussed in \citet[][]{Hon2022}, for a uniform stellar population within the spheroidal component, there should be no systematic change in the spheroid's size with wavelength, modulo the impact of dust. 
We are, however, at present unable to quantify this change from the $i$-band to 3.6 microns, and as such, have not implemented any rescaling of the spheroid sizes.

The correlation coefficients for the entire sample, $r_\mathrm{p}(\rm All)$ and $r_\mathrm{s}(\rm All)$, where "All" now includes all four galaxy samples (SG16, D+19, S+19, and H+22) in both $\log( M_{\rm *,Sph})$--$\log(n_\mathrm{Sph})$, $\log( M_{\rm *,Sph})$--$\log(\Sigma_\mathrm{0,Sph})$, and $\log( M_{\rm *,Sph})$--$\log(R_\mathrm{e,Sph})$ planes are comparable to that in the $\mathfrak{M}_{\rm Sph}$--$\log(n_\mathrm{Sph})$, $\mathfrak{M}_{\rm Sph}$--$\log(\mu_\mathrm{0,Sph})$, and $\mathfrak{M}_{\rm Sph}$--$\log(R_\mathrm{e,Sph})$ planes from Fig.~\ref{fig:M_n_mu}, respectively.
The inclusion of the data from SG16, D+19 and S+19 does not increase the correlation strength.
In the $\log(M_{\rm *,Sph})$--$\log(\Sigma_\mathrm{0,Sph})$ plane, when we divide the sample into ellipticals and disc-galaxies, both subgroups return an extremely low correlation, with $r_\mathrm{p}( E+ES) = 0.19$ and $r_\mathrm{p}(S0+S) = 0.44$. 
The weak correlations between the spheroid luminosity (subsequently its stellar mass) and projected central density persists even after being converted into physical quantities, $M_*$ and $\Sigma_\mathrm{0,Sph}$. 
\citet{Sahu2022} reported a similar Pearson correlation coefficient, $r_\mathrm{p}(All)=0.57$ (their table 1), in the $M_\mathrm{BH}$--$\Sigma_\mathrm{0,Sph}$ plane. 
Due to the coevolution between the SMBH and the spheroid, $M_\mathrm{BH}$ and $M_\mathrm{Sph}$ are strongly correlated.
Therefore, the similarity between the $\log(M_\mathrm{BH})$--$\log(\Sigma_\mathrm{0,Sph})$ and $\log(M_\mathrm{*,Sph})$--$\log(\Sigma_\mathrm{0,Sph})$ relation is somewhat expected.

In Fig.~\ref{fig:Rz_sigmaz}, we convert the $\log(M_\mathrm{*,Sph})$--$\log(R_\mathrm{e,Sph})$ and $\log(M_\mathrm{*,Sph})$--$\log(\Sigma_\mathrm{0,Sph})$ into the $\log(M_\mathrm{*,Sph})$--$\log(R_{z,\rm Sph})$ and $\log(M_\mathrm{*,Sph})$--$\log(\Sigma_{z,\rm Sph})$ distributions by using different scale radii, encompassing light fractions $z= 0.05, 0.5,$ and $0.95$.
The behaviour is similar to that in Fig.~\ref{fig:diff_R}.
The $\log(M_\mathrm{*,Sph})$--$\log(R_{z,\rm Sph})$ distributions maintain a somewhat linear trend with positive slope while the $\log(M_\mathrm{*,Sph})$--$\log(\Sigma_{z,\rm Sph})$ distributions change slope with different values of $z$.

\subsubsection{\texorpdfstring{$M_{\rm *,Sph}$--$\langle\Sigma\rangle_{z,\rm Sph}$}{} relation}
\label{sec:average_sigma0}

Some studies have suggested that the average projected mass density within a small fixed radius, say 1 kpc, $\langle \Sigma \rangle_\mathrm{1~kpc}$, or 5 kpcs, $\langle \Sigma \rangle_\mathrm{5~kpc}$, are great estimators to track the growth of SMBHs \citep[e.g.,][]{Barro2017, Dekel2019, Ni2021, Sahu2022}.
By proxy, it will also pertain to the growth of spheroids, given the connection between the two. 
Indeed, the $M_\mathrm{BH}$--$\langle \Sigma \rangle_\mathrm{5~kpc, Sph}$ plane exhibits a log-linear relation \citep[$r_\mathrm{p} = 0.83$ and $r_\mathrm{s} = 0.84$, see][]{Sahu2022}.
However, it is noteworthy that at this fixed radius, the quantity $\langle \Sigma \rangle_\mathrm{5~kpc, Sph}$ might be sampling a different percentage of light in different spheroids.
For instance, for a bulge with $R_\mathrm{e} = 6~\rm kpc$, $\langle \Sigma \rangle_\mathrm{5~kpc, Sph}$ captures nearly half of the light, while for a spheroid like an E galaxy with $R_\mathrm{e} = 12~\rm kpc$, it captures a small fraction.
In the interest of studying how the average projected mass density\footnote{$\langle\Sigma\rangle_{z,\rm Sph}$ can be calculated by subsituting $\langle\mu\rangle_{z,\rm Sph}$ (see equation~\ref{eq:average_muz}) into equation~\ref{eq:sigma0_law}.}, $\langle\Sigma\rangle_{z,\rm Sph}$, varies as one includes a different percentage of light within the radius $R_{z,\rm Sph}$, we plot $M_\mathrm{*, Sph}$ versus $\langle \Sigma \rangle_{z,\rm Sph}$ in Fig.~\ref{fig:average_sigma}.
The $\log(M_\mathrm{*,Sph})$--$\log(\langle \Sigma \rangle_{z,\rm Sph})$ relations present a negative trend for all $z$. 
This result is consistent with \citet{Sahu2022} who found a negative correlation in the $\log(M_\mathrm{BH})$--$\log(\langle \Sigma \rangle_\mathrm{e, Sph})$ ($=\log(\langle \Sigma \rangle_\mathrm{0.5, Sph})$) relation (see their Fig.~6).
As the value of $z$ increases, the $\log(M_\mathrm{*, Sph})$--$\log(\langle \Sigma \rangle_{z, \rm Sph})$ relation appears increasingly more linear and has a steeper slope.
In addition, it appears that the correlation strength of the distributions varies with $z$.
At $z=0.05$, the correlation is non-existent ($r_p, r_s \sim -0.08$) while at $z=0.95$, $\log(M_\mathrm{*,Sph})$ and $\log(\langle\Sigma\rangle_{z, \rm Sph})$ have a moderate correlation ($r_p, r_s \sim -0.68$).

\begin{figure*} 
\centering
	\includegraphics[ width=0.95\textwidth]{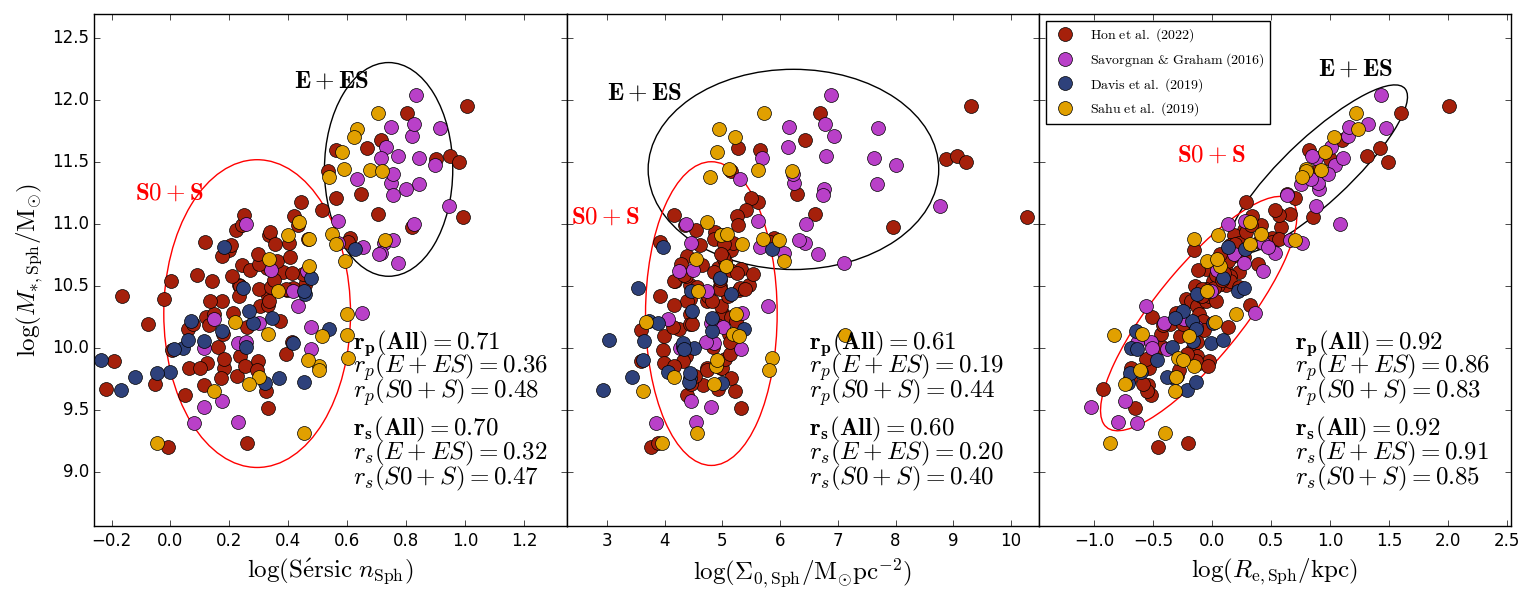}
    \caption{Left-hand panel: Spheroid mass versus S\'{e}rsic index. 
    Middle panel: Spheroid mass versus projected surface mass density, $\Sigma_\mathrm{0, Sph}$. 
    Right-hand panel: Spheroid mass versus effective radius, $R_\mathrm{e,Sph}$.
    Similar to Fig.~\ref{fig:M_n_mu}, the correlation coefficients are also shown.
    The points are colour-schemed according to the data sources.
    The black and red ellipses are drawn to highlight E+ES galaxies and S0+S bulges, respectively.
    They represent a $2\sigma$ range about the median value of the two subgroups.
    }
    \label{fig:mass_sigma0}
\end{figure*}

\begin{figure*} 
\centering
	\includegraphics[width=0.8\textwidth]{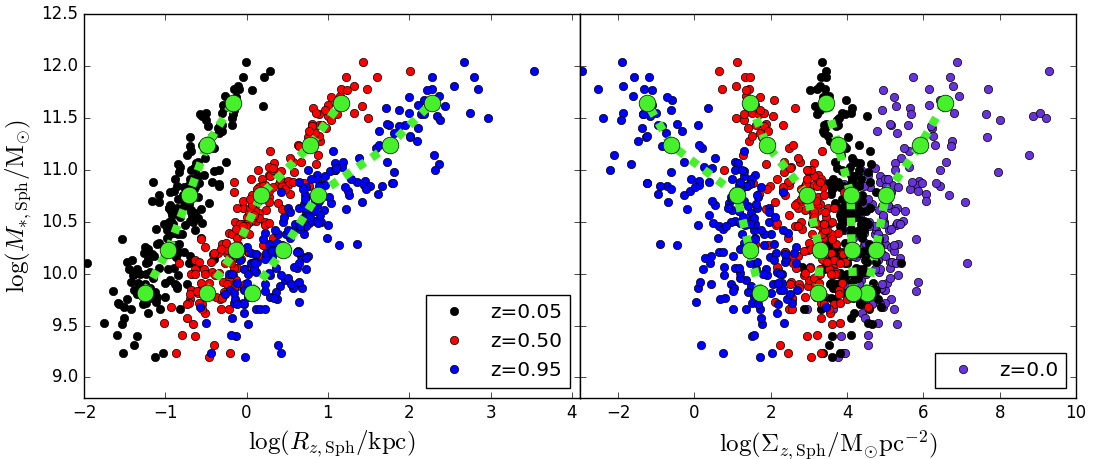}
    \caption{
    Left-hand panel: Spheroid mass versus scale radii $R_{z, \rm Sph}$.
    Right-hand panel: Spheroid mass versus projected stellar mass density, $\Sigma_{z, \rm Sph}$.
    Three fractional values of $z = 0.05, 0.50, 0.95$, are shown in different colours.
    The data points (purple) in the right-hand panel with $z=0$ are the same as the ones shown in the middle panel of Figure~\ref{fig:mass_sigma0}.}
    \label{fig:Rz_sigmaz}
\end{figure*}

\begin{figure} 
\centering
	\includegraphics[width=0.45\textwidth]{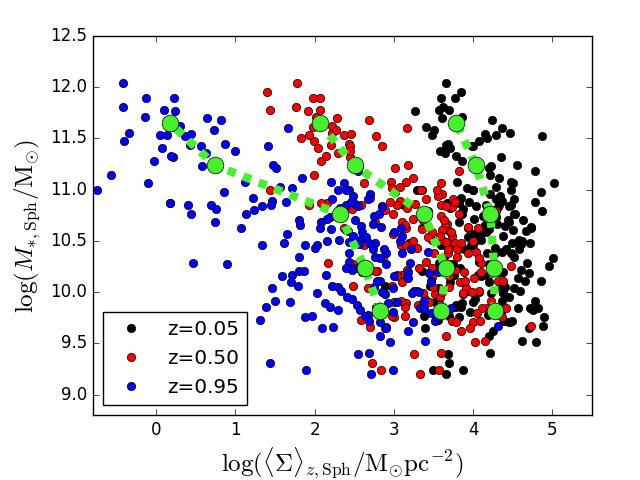}
    \caption{Spheroid mass versus the average projected stellar mass density, $\langle \Sigma\rangle_{z, \rm Sph}$.}
    \label{fig:average_sigma}
\end{figure}

\begin{figure} 
\centering
	\includegraphics[width=0.47\textwidth]{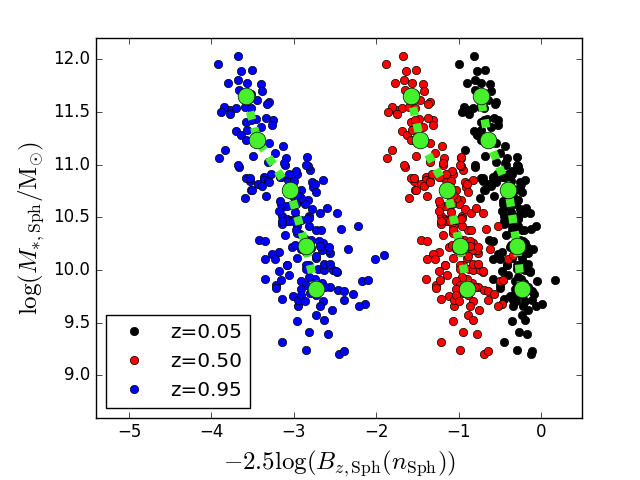}
\caption{The spheroid mass, $\log(M_\mathrm{*,Sph})$, versus $-2.5\log(B_{z, \rm Sph}(n_{\rm Sph}))$ (see equation~\ref{eq:average_Bnz}). 
Similar to Fig.~\ref{fig:diff_R}, \ref{fig:Rz_sigmaz}, and \ref{fig:average_sigma}, the sample is divided by different percentage of light, $z$, captured in $R_{z, \rm Sph}$.}
    \label{fig:term_b}
\end{figure}

Since $\langle \Sigma\rangle_{z,\rm Sph}$ is calculated by substituting $\mu_{z,\rm Sph}$ with $\langle \mu\rangle_{z,\rm Sph}$ in equation~\ref{eq:sigma0_law}, the varying trends in the $\log(M_\mathrm{*,Sph})$--$\log(\langle\Sigma\rangle_{z, \rm Sph})$ plane can be explained by looking into the formulation in equation~\ref{eq:average_muz}.
The average surface brightness, $\langle \mu\rangle_{z,\rm Sph}$, is a linear combination of the surface brightness, $\mu_z$, in the first term and the shape function, $B_z(n)$, in the second term.
Because $B_{z,\rm Sph}(n)$ is purely a function of S\'ersic index $n$, it inherits the moderately strong linearity in the $\log(M_\mathrm{*,Sph})$--$\log(n_\mathrm{Sph})$ plane (see Fig.~\ref{fig:mass_sigma0}).
We demonstrate such a strong trend in Fig.~\ref{fig:term_b}, where the spheroid stellar mass ($M_\mathrm{*,Sph}$) is plotted against $-2.5~\log[B_{z, \rm Sph}(n_{\rm Sph})]$. 
Indeed, the $\log(M_\mathrm{*,Sph})$--$(-2.5\log[B_{z, \rm Sph}])$ relations have a rather consistent anti-correlation\footnote{The negative sign is due to the multiplier $-2.5$. Spheroid stellar mass, $\log(M_\mathrm{*,Sph})$ correlates positively with the shape function $\log(B_{z,\rm Sph})$.} across $z$, with $r_p, r_s\sim -0.7$.
Hence, the variation in correlation strength in the $\log(M_\mathrm{*,Sph})$--$\log(\langle\Sigma\rangle_{z,\rm Sph})$ plane is because of the inherent non-linearity in the $\log(M_\mathrm{*, Sph})$--$\rm \mu_{z, \rm Sph}$ relation (see Fig.~\ref{fig:M_n_mu}).
This latter trend dampens the correlation strength of the $\log(M_\mathrm{*,Sph})$--$(-2.5\log[B_{z, \rm Sph}])$ distribution and results in an changed trend as $z$ changes.
It is clear that the choice of the light fraction value $z$ significantly affects the shape and slope of the scaling relations. 

\subsubsection{Correlation strength ($r_\mathrm{p}$ \& $r_\mathrm{s}$) versus fraction $z$}
\label{sec:corr_trend}

\begin{figure} 
\centering
	\includegraphics[width=0.5\textwidth]{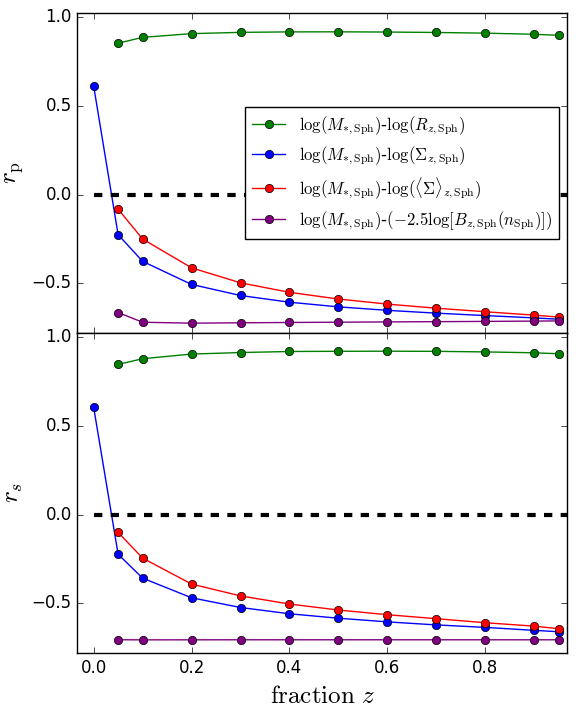}
    \caption{The correlation strength of different quantities versus the spheroid stellar mass as a function of the fractional light $z$. 
    The upper and lower panel shows the Pearson and Spearman correlation coefficient, $r_p$ and $r_s$, respectively.  
    Both plots shows the correlation coefficient for the following relations:  $\log(M_\mathrm{*,Sph})$--$\log(R_{z, \rm Sph})$ (green),
    $\log(M_\mathrm{*,Sph})$--$\log(\Sigma_{z, \rm Sph})$ (blue),
    $\log(M_\mathrm{*,Sph})$--$\log(\langle\Sigma\rangle_{z,\rm Sph})$ (red), and $\log(M_\mathrm{*,Sph})$--$(-2.5\log[B_{z, \rm Sph}(n_{\rm Sph})])$ (purple).}
    \label{fig:corr_trend}
\end{figure}

Here, we explore how the correlation strength changes with different light fractions $z$.
In Fig.~\ref{fig:corr_trend}, we plot the correlation strength, $r_p$ and $r_s$, in the upper and lower panel, respectively, as a function of the fraction of spheroid light, $z$, contained within the radius $R_{z, \rm Sph}$, for various scaling relations: $\log(M_\mathrm{*,Sph})$--$\log(R_{z, \rm Sph})$,  $\log(M_\mathrm{*,Sph})$--$\log(\Sigma_{z, \rm Sph})$, $\log(M_\mathrm{*,Sph})$--$\log(\langle\Sigma\rangle_{z, \rm Sph})$, and $\log(M_\mathrm{*,Sph})$--$(-2.5\log[B_{z. \rm Sph}(n)])$, where all involve the stellar mass.
The $\log(M_\mathrm{*,Sph})$--$\log(R_{z, \rm Sph})$ relations (green line) are consistently strongly coupled regardless of the value of $z$.
The $\log(M_\mathrm{*,Sph})$--$\log(\Sigma_{z, \rm Sph})$ relation (blue line) is positively correlated at $z=0$, but it becomes negatively correlated at larger $z$, a feature that can be seen in the right-hand panel of Fig.~\ref{fig:Rz_sigmaz}.
For the $\log(M_\mathrm{*,Sph})$--$\log(\langle\Sigma\rangle_{z, \rm Sph})$ relation at $z = 0.05$ (first point on the red line), it has an extremely low correlation ($r_\mathrm{p}, r_\mathrm{s}\sim0$) but as $z$ increases, it becomes increasingly anti-correlated.
Finally, the $\log(M_\mathrm{*,Sph})$--$(-2.5\log[B_{z, \rm Sph}(n_{\rm Sph})])$ relations (purple line) exhibit a very consistent anti-correlation trend across all $z$ with $r_{p}, r_{s} \sim -0.7$.

Fig.~\ref{fig:corr_trend} provides an important insight into spheroid scaling relations that, in general, the surface brightness, $\mu_{z, \rm Sph}$, and its derived physical quantities ($\Sigma_{z, \rm Sph}$ and $\langle \Sigma \rangle_{z, \rm Sph}$) are comparatively weak estimators for spheroid stellar mass.
While the scaling relations become more reliable at higher light fraction $z$, the correlation coefficients are limited to $\sim0.7$.
The variability in correlation strength across $z$ makes these scaling relations inherently biased by choice of $R_{z, \rm Sph}$.

In Section~\ref{sec:Struc_para}, we found that the S\'ersic index ($n_\mathrm{Sph}$) of a spheroid is a slightly better predictor for a spheroid stellar mass than $\mu_0$.
Here, we show that the S\'ersic shape function, $B_{z, \rm Sph}(n_{\rm Sph})$, is also a decent spheroid mass predictor across all $z$ ($r_p, r_s \sim 0.7$).
Similarly, $R_{z,\rm Sph}$ also has a consistently strong log-linear relation with $M_\mathrm{*,Sph}$ across all $z$ but with an even higher correlation strength ($r_p, r_s \sim 0.9$).
Hence, we advocate using $R_{z,\rm Sph}$ as the primary predictor to spheroid stellar mass ($M_\mathrm{*,Sph}$) and $n_\mathrm{Sph}$ (or $B_{z, \rm Sph}(n_{\rm Sph})$) as the secondary.
Of course, $R_{z, \rm Sph}$, $n_\mathrm{Sph}$, and $\mu_{0, \rm Sph}$ perfectly define the stellar luminosity of the spheroid.

\subsection{Fitting the scaling relations}

\subsubsection{The spheroid size--mass (\texorpdfstring{$R_\mathrm{e,Sph}$}{}--\texorpdfstring{$M_\mathrm{*,Sph}$}{}) scaling relation} 
\label{sec:size_mass}

\begin{figure*} 
\centering
	\includegraphics[ width=0.45\textwidth]{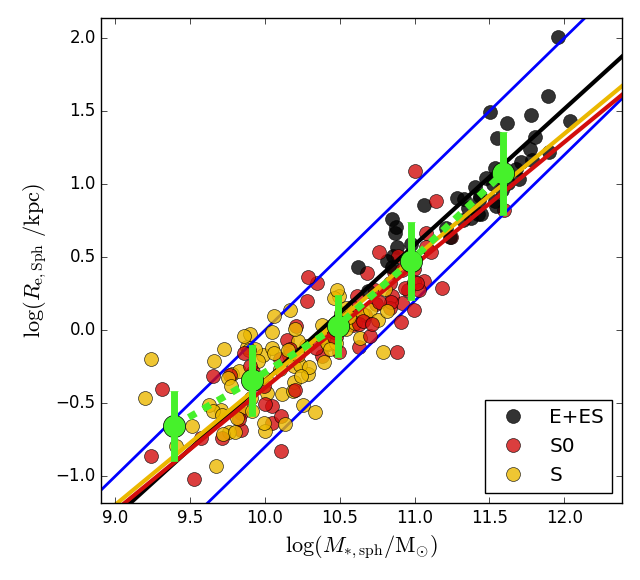}
	\includegraphics[ width=0.45\textwidth]{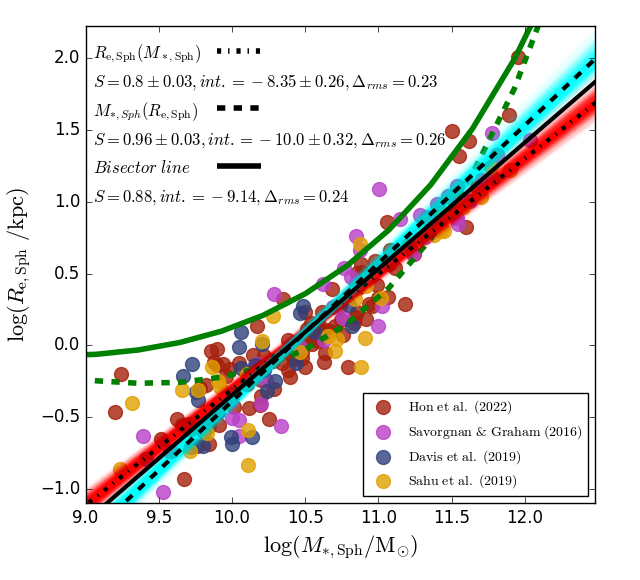}
    \caption{Left-hand panel: Size--mass ($\log(R_\mathrm{e,Sph})$--$\log(M_\mathrm{*,Sph})$) relation for 202 local spheroids with $z\lesssim0.05$ divided by the host galaxies' morphology from the following works involving multicomponents decomposition: SG16, D+19, S+19, and H+22.
    The size is from the geometric-mean radius.
    The sample is separated by elliptical (E) + ellicular (ES) galaxies (black points), spheroids embedded in lenticular (S0) galaxies (red points) and spiral (S) galaxies (orange points). 
    The light green dashed track is made by binning the data points in between a mass interval of 0.6 dex across $M_*\sim 10^{9}$--$10^{12}~\rm M_{\odot}$.
    The green points on the dashed track are the median values, and the error bars indicate the standard deviations $\pm1\sigma$ for each interval. 
    The two blue lines are two single-power laws (equation~\ref{eq:power_law}) with $S=1.0$ and $\rm int. = -10.0~\& -10.8$.
    The colour-coded lines are obtained via fitting a single power law to the spheroids in each morphology group via bisector regression.
   Right-hand panel: Same data points separated by their sources: SG16 (purple points), D+19 (blue points), S+19 (orange points), and H+22 (red points).
   The data points are fitted by a log-linear relation (see equation~\ref{eq:power_law}).
    The red shaded area consists of 5000 MCMC trial fits when treating the stellar mass, $M_\mathrm{*, Sph}$, as the independent variables.
    The dash-dotted black line is the optimal fit among the trials.
    The cyan shaded area and the dashed black line are similar but with the effective radius ($R_\mathrm{e, Sph}$) treated as the independent variable.
    The solid black line is the bisector line constructed using the preceding two approaches.
    The effective range for our relation is within $2\times 10^{9}\lesssim M_\mathrm{*,Sph} / \rm M_{\odot} \lesssim 2\times10^{12}$.
    For comparison, the \textit{curved} size-mass relation for ETGs in the $B$-band modelled with a single S\'ersic function \citep{Graham2006A} and the bulges obtained from S\'ersic+exponential models \citep[][]{Graham2008} in the $K$-band are plotted as solid and dashed dark green lines, respectively. 
    To compare the bulges with the ETGs, we adjust their bulge magnitude by $B-K = 4.0~\rm mag$ to conform with the $B$-band magnitude.
    Their magnitudes are converted to stellar mass using a constant mass-to-light ratio of $M/L_B= 3.9$ (see texts in Section~\ref{sec:comparison} for details).}
    \label{fig:size_mass_morphology}
\end{figure*}

With the above observation that spheroid size ($R_{z, \rm Sph}$) is the most consistent and reliable predictor for spheroid stellar mass ($M_\mathrm{*, Sph}$) across all $z$, we proceed to investigate the spheroid size-mass relation for our expanded sample. 
In Fig.~\ref{fig:size_mass_morphology}, we present the equivalent axis effective radius ($R_\mathrm{e,Sph}$) versus the stellar mass ($M_\mathrm{*,Sph}$) for the local spheroids from SG16, D+19, S+19, and H+22. 
In the left-hand panel of Fig.~\ref{fig:size_mass_morphology}, the data is separated by the morphological type of the host galaxies.
There is a significant overlap between S0 and S spheroids.
The elliptical (E) and ellicular (ES) galaxies are the largest and most massive, occupying the region $R_\mathrm{e,Sph} \gtrsim 2~\rm kpc$ and $M_\mathrm{*,Sph} \gtrsim 10^{11}~\rm M_{\odot}$.
They are also more likely to have a depleted stellar core: 19 of them are well described by the core-S\'{e}rsic model.
Spheroids embedded in S0 and S galaxies reside within the range $0.2~\lesssim R_\mathrm{e,Sph}/\rm kpc \lesssim ~5$ and $3 \times 10^{9} \lesssim~M_\mathrm{*,Sph}/\rm M_{\odot}~\lesssim 2\times10^{11}$.

In the right-hand panel of Fig.~\ref{fig:size_mass_morphology}, the data points are colour-coded for their respective sources.
The size--mass relation shows strong agreement between the four data sets despite having different personnel conducting the decomposition. 
The big spheroids follow an established trend set by elliptical galaxies (not to be confused with ETGs).
Although there is an important upturn at high masses, which we speculate is due to E+E mergers rather than E built from S0+S0 mergers, we fit a single power-law to the relation using the \texttt{Linmix} fitting routine \citep[][]{Kelly2007}:

\begin{equation} 
\label{eq:power_law}
    \log(R_\mathrm{e}) = \mathrm{S} \log\bigg( M_{*}\bigg)+ \mathrm{int}.\\
\end{equation}
where $S$ is the slope and $int.$ is the y-intercept of the relation.

\texttt{Linmix} calculates the best-fit regression line via maximising the Bayesian likelihood function  \citep[see equation~16 in][]{Kelly2007}.
The prior distribution of the independent variable is assumed to be a Gaussian mixture.
Such treatment allows greater flexibility in dealing with data heteroscedasticity, i.e., the non-uniform variance in each data point.
We implemented 5000 iterations of Monte Carlo Markov Chain (MCMC) sampling to maximise the likelihood function. 
It is unclear in nature whether the size of a galactic structure depends on its stellar mass or vice versa.
For this reason, we have performed the fitting process with $M_\mathrm{*, Sph}$ treated as the independent variable for the relation $R_\mathrm{e,Sph}(M_\mathrm{*, Sph})$ and $R_\mathrm{e,Sph}$, as the independent variable for the relation $M_\mathrm{*, Sph}(R_\mathrm{e,Sph})$.
In the right-hand panel of Fig.~\ref{fig:size_mass_morphology}, we depict the regression lines for fitting $R_\mathrm{e,Sph}(M_\mathrm{*, Sph})$ and $M_\mathrm{*, Sph}(R_\mathrm{e,Sph})$ with red and cyan shaded area, respectively.
The optimal fits among these regressions are shown as a dash-dotted and dashed line, respectively.
They are the following relations.
\begin{subequations} 
\label{eq:size_mass_twofit_morph}
\begin{align*}
 R_\mathrm{e,Sph}(M_\mathrm{*,Sph}):& \\
 \log(R_\mathrm{e,Sph}/\rm kpc)&=(0.80\pm0.03)\log(M_\mathrm{*,Sph}/\rm M_{\odot})-(8.35\pm0.26), \tag{12a}\\
 M_\mathrm{*,Sph}(R_\mathrm{e,Sph}):& \\ \log(R_\mathrm{e,Sph}/\rm kpc)&=(0.96\pm0.03)\log(M_\mathrm{*,Sph}/\rm M_{\odot})-(10.00\pm0.32), \tag{12b}
\end{align*}
\end{subequations}
From these regression lines, we calculated the bisector line, shown as the solid black line in the panel:
\begin{subequations} 
\label{eq:size_mass_bisector}
\begin{align*}
 \log(R_\mathrm{e,Sph}/\rm kpc)&=0.88\log(M_\mathrm{*,Sph}/\rm M_{\odot})-9.15, \tag{13}
\end{align*}
\end{subequations}
with a scatter of $\Delta_{rms} = 0.24~\rm dex$ in the vertical direction.
All three scaling relations are valid but with differing assumptions.
Future work may choose the appropriate relation for their purpose.
However, since the spheroid size and mass are highly correlated ($r_p\sim0.9$), the three scaling relations are similar.

For $R_\mathrm{e} \lesssim 10~\rm kpc$, a log-linear relation appears adequate to describe the size-mass distribution of local spheroids. 
An unrelated log-linear relation was also demonstrated in \citet[][see their Fig.~8]{Sahu2020} but is now abundantly clear with the additional 103 spheroids from H+22.
As noted above, there is a slight departure from a log-linear relation at the high-mass end ($M_\mathrm{*, Sph} > 5\times 10^{11}~\rm M_{\odot}$), where, at fixed mass, massive spheroids have a larger radius than the log-linear relation suggests. 
We speculate that this might also partly be due to the influence of intracluster light (ICL) in the brightest cluster galaxies (BCGs)\footnote{There are three, four, and three BCGs in SG16, S+19, and H+22, respectively. 
All of which have a galaxy stellar mass $M_\mathrm{*,gal} > 4 \times10^{11}~\rm M_{\odot}$.}.
The most massive ($M_* \gtrsim 4 \times10^{11}~\rm M_{\odot}$) galaxies tend to be BCGs living in the centre of clusters with extended stellar halos.
As a result, the ICL accumulates around the BCG. 
When modelling the galaxies with a single S\'ersic model, we could be slightly biased by the ICL and overestimate their intrinsic size.
For E+E mergers in which the velocity dispersion, $\sigma$, may not increase, the virial theorem ($M_{*} \propto \sigma^{2}R$) dictates $M_{*} \propto R$.
For reference, we have added solid blue lines of slope $S=1$ in Fig.~\ref{fig:size_mass_morphology}.

\subsubsection{Additional scaling relations involving S\'ersic index}
\label{sec:additional_relations}

We present a few supplementary scaling relations: $\log(n_\mathrm{Sph})$--$\log(M_\mathrm{*,Sph})$ and $\log(B_\mathrm{e,Sph})$--$\log(M_\mathrm{*,Sph})$, $\log(\Sigma_\mathrm{0,Sph})$--$n_\mathrm{Sph}$, and $\log(R_\mathrm{e,Sph})$--$\log(n_\mathrm{Sph}$).
These extra relations have a moderate to high linearity ($r_p \approx 0.7-0.9$) and, therefore, are suited to be secondary estimators for spheroid structures. 
Our intention here is to merely demonstrate and touch on the potential importance of these relations.
A deeper exploration of their physical meaning shall be conducted in future works.

\begin{figure*} 
\centering
	\includegraphics[width=0.46\textwidth]{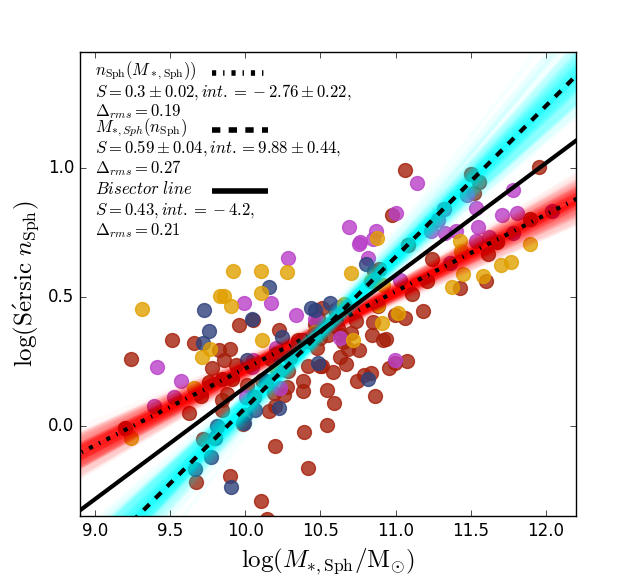}
	\includegraphics[width=0.46\textwidth]{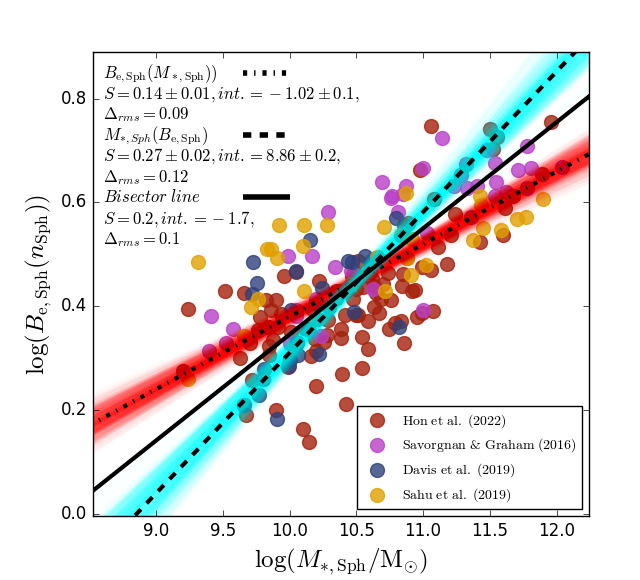}
    \caption{Left-hand panel: Spheroid S\'ersic index-stellar mass ($\log(n_\mathrm{Sph})$--$\log(M_\mathrm{*, Sph})$) relation. 
    Right-hand panel: Spheroid S\'ersic shape function (at light fraction $z=0.5$)--versus--stellar mass ($\log(B_\mathrm{e,Sph})$--$\log(M_\mathrm{*, Sph})$) relation. 
    In both panels, the red shaded area depicts the 5000 MCMC trial fits when treating $M_\mathrm{*, Sph}$ as the independent variable. 
    The black dash-dotted line is the optimal fit among these trials.
    The cyan shaded area and black dashed lines are similar but with  $n_\mathrm{Sph}$ (or $B_\mathrm{e, Sph}$) treated as the independent variable during the fitting process.
    The solid black line is the bisector line obtained from the two optimal fits mentioned above.}
\label{fig:Be_mass_fit_coparison}
\end{figure*}

S\'ersic index, $n$, and its derived `shape function', $B_z(n)$, have a weaker correlation with stellar mass than the size--mass relation, with $r_p, r_s \sim 0.72$ versus $\sim0.9$.
The left- and right-hand panel of Fig.~\ref{fig:Be_mass_fit_coparison} depict the $\log(n_{\rm Sph})$--$\log(M_\mathrm{*,Sph})$ plane and the $\log[B_\mathrm{e,Sph}(n_{\rm Sph})]$--$\log(M_\mathrm{*,Sph})$ plane\footnote{$B_z(n)$ (equation~\ref{eq:average_Bnz}) depends on $b_{n,z}$ and thus $R_z$, Here, we set $R_z$ to $R_\mathrm{e}$, i.e. $z=0.5$.}, respectively.
We performed multiple linear fits on our sample, similar to what was done for the size--mass relation in Section~\ref{sec:size_mass}.
For the $\log(n_\mathrm{Sph})$--$\log(M_\mathrm{*,Sph})$ relation, we have the following bisector relations:
\begin{subequations} 
\label{eq:n_mass_twofit}
\begin{align*}
 \log(n_\mathrm{Sph})&=0.43\log(M_\mathrm{*,Sph}/\rm M_{\odot})-4.20, \tag{14}
\end{align*}
\end{subequations}
with a scatter of $\Delta_{rms}=0.21~\rm dex$.

For the $\log(B_\mathrm{e,Sph})$--$\log(M_\mathrm{*,Sph})$ relation, we have:
\begin{subequations} 
\label{eq:Bn_mass_twofit}
\begin{align*}
 \log(B_\mathrm{e,Sph})&=0.20\log(M_\mathrm{*,Sph}/\rm M_{\odot})-1.70. \tag{15}
\end{align*}
\end{subequations}
with a scatter of $\Delta_{rms}=0.10$ dex.

The $\log(B_{z, \rm Sph}(n_{\rm Sph}))$--$\log(M_\mathrm{*,Sph})$ scaling relations also present a smaller scatter than the $\log(n_{\rm Sph})$--$\log(M_\mathrm{*,Sph})$ scaling relation.
Note that this is due to the S\'ersic shape function, $B_{z, \rm Sph}(n_\mathrm{Sph})$, re-scaled the input parameter, $n_\mathrm{Sph}$ into a smaller range and, therefore, created a smaller scatter in the vertical axis of the plot.

\begin{figure} 
\centering
	\includegraphics[width=0.46\textwidth]{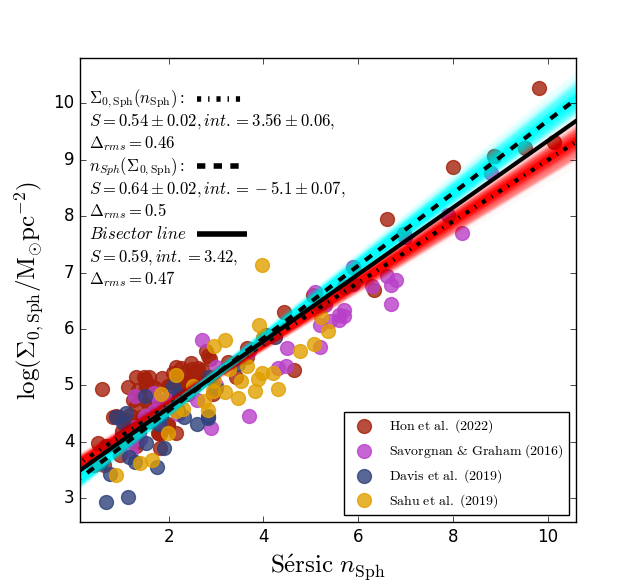}
    \caption{
    The spheroid central projected mass density--versus--S\'ersic index ($\log(\Sigma_\mathrm{0, Sph})$--$n_\mathrm{Sph}$) relations.}
\label{fig:mu_n_fit}
\end{figure}

We plotted the bisector fit in the $\log(\Sigma_\mathrm{0,Sph})$--$n_\mathrm{Sph}$ plane in Fig.~\ref{fig:mu_n_fit}.
We have obtained the following scaling relations:
\begin{subequations} 
\label{eq:sigma0_n_twofit}
\begin{align*}
 \log(\Sigma_\mathrm{0,Sph}/\rm M_{\odot}pc^{-2})&=0.59~n_\mathrm{Sph}+3.42. \tag{16}
\end{align*}
\end{subequations}
with a scatter of $\Delta_{rms}=0.47~\rm dex$.
Crucially, $\log(\Sigma_\mathrm{0,Sph})$ present a strong linear relation against $n_\mathrm{Sph}$ instead of $\log(n_\mathrm{Sph})$.

\begin{figure} 
\centering
	\includegraphics[ width=0.47\textwidth]{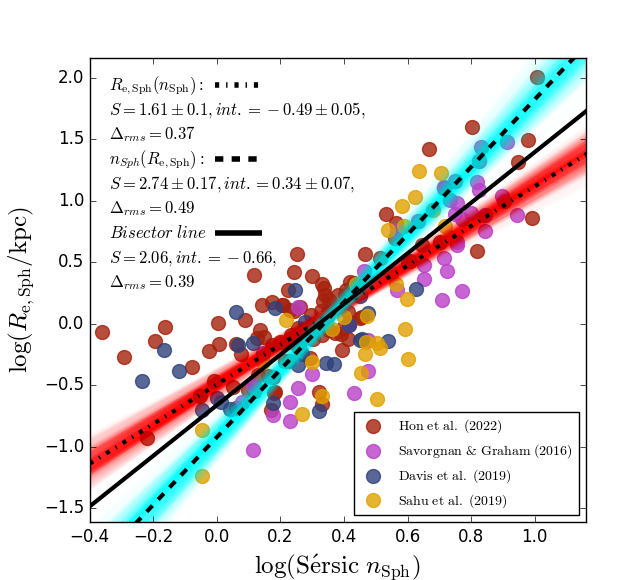}
    \caption{The spheroid size--versus--S\'ersic index ($\log(R_\mathrm{e,Sph})$--$\log(n_\mathrm{Sph})$) relation.}
\label{fig:Re_n_fit}
\end{figure}

Previously, we found that two pairs of quantities are strongly coupled in spheroids: the size ($R_\mathrm{e,Sph}$)--mass ($M_\mathrm{*,Sph}$), and the surface brightness ($\mu_\mathrm{0,Sph}$)--shape parameters ($n_\mathrm{Sph}$).
One may question if the size ($R_\mathrm{e,Sph}$) and shape ($n_\mathrm{Sph}$) are also somehow related.
From Section~\ref{sec:Struc_para}, we know that the correlation strength in the $\log(R_\mathrm{e,Sph})$--$\log(n_\mathrm{Sph})$ plane for the H+22 sample is moderately strong ($r_p = 0.79$ and $r_s = 0.73$).
In Fig.~\ref{fig:Re_n_fit}, we performed a symmetrical regression on the spheroid in the $\log(R_\mathrm{e,Sph})$--$\log(n_\mathrm{Sph})$ plane. 
The bisector scaling relation is:
\begin{subequations} 
\label{eq:R_n_twofit}
\begin{align*}
 \log(R_\mathrm{e,Sph}/\rm kpc)&=2.06\log(n_\mathrm{Sph})-0.66, \tag{17}
\end{align*}
\end{subequations}
with a scatter of $\Delta_{rms}=0.39~\rm dex$.

It is curious that the two pairs of quantities, $\log(R_\mathrm{e, Sph})$--$\log(M_\mathrm{*,Sph})$ and $\log(\Sigma_\mathrm{0,Sph})$--$n_\mathrm{Sph}$, have such a strong correlation ($r_p, r_s \sim 0.9$) while $\log(n_\mathrm{Sph})$--$\log(M_\mathrm{*,Sph})$ and $\log(R_\mathrm{e,Sph})$--$\log(n_\mathrm{Sph})$ do not ($r_p, r_s \sim 0.7$).
The $\log(R_\mathrm{e,Sph})$--$\log(n_\mathrm{Sph})$ relation therefore appears a secondary scaling relations that connects the two primary scaling relations: $\log(R_\mathrm{e,Sph})$--$\log(M_\mathrm{*,Sph})$ and $\log(\Sigma_\mathrm{0,Sph})$--$\log(n_\mathrm{Sph})$. 

If the `photometric plane\footnote{An unifying plane in the parameter space of  ($\log(n),\mu_0, \log(R_\mathrm{e})$).}' \citep[][]{Khosroshahi2000B} exist for spheroids, we might be seeing a side of the plane's surface in the $\log(R_\mathrm{e})$--$\log(n_\mathrm{Sph})$ distribution.
Combining with the knowledge that, unlike in \citet[][]{Khosroshahi2000B}, our $\log(\Sigma_\mathrm{0,Sph})$--$n_\mathrm{Sph}$ relation appears to be linear instead of \textit{curved},
the photometric plane for our spheroids might be a curved surface.
Further investigation, however, is beyond the scope of this paper and shall be explored in future works.

\section{Discussion and Conclusions} 
\label{sec:Discussion}

\begin{figure*} 
\centering
	\includegraphics[ width=1.0\textwidth]{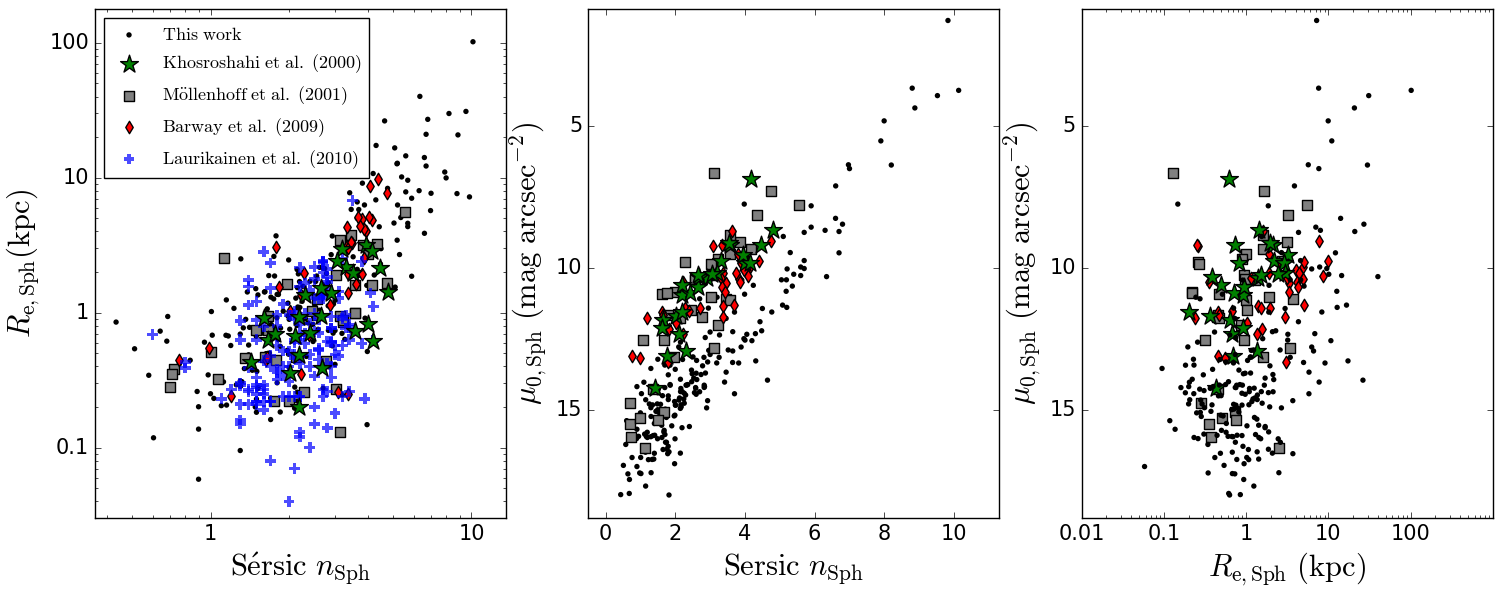}
    \caption{
    A comparison of the spheroid parameters with some early-works.
    Left-hand panel: The $\log(R_\mathrm{e,Sph})$--$\log(n_\mathrm{Sph})$ plane.
    Middle panel: The $\mu_\mathrm{0,Sph}$--$n_\mathrm{Sph}$ plane.:   
    Right-hand panel: The $\mu_\mathrm{0,Sph}$--$\log(R_\mathrm{e,Sph})$ plane.
    In all three panels, we present the bulge data from \citet[][green $\star$]{Khosroshahi2000A}, \citet[][grey $\blacksquare$]{Mollenhoff2001}, \citet[][red $\blacklozenge$]{Barway2009}, and \citet[][blue +]{Laurikainen2010}. See text for a description of the individual studies and the adjustment for differing $\mu_0$ bandpass ($i-K = 2.48$).
    }
\label{fig:3plot_compare}
\end{figure*}

\begin{figure} 
\centering
	\includegraphics[ width=0.45\textwidth]{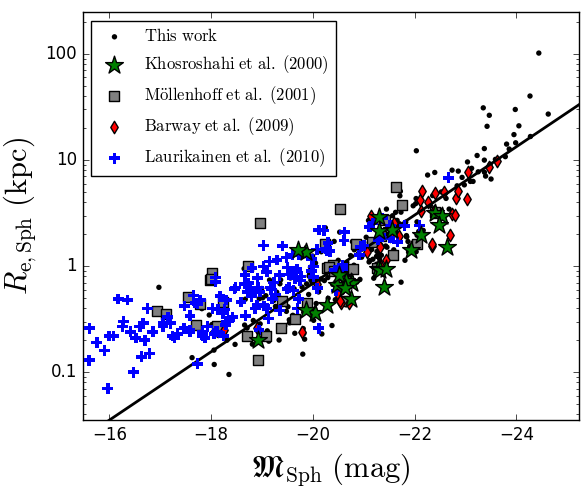}
    \caption{Comparison with works from the literature: bulge size--versus--absolute magnitude ($R_\mathrm{e, Sph}$--$\mathfrak{M}_\mathrm{i, Sph}$). 
    The symbols and colour follows Fig.~\ref{fig:3plot_compare}.
    The $K$-band are artificially shifted to match with our spheroid sample by $\Delta \mathfrak{M} = +2.48~\rm mag$ for comparison.
    The solid black line is an ordinary least square fit on our data points with a log-linear function.
    }
\label{fig:Re_Mag_fit}
\end{figure}

\begin{figure} 
\centering
	\includegraphics[ width=0.47\textwidth]{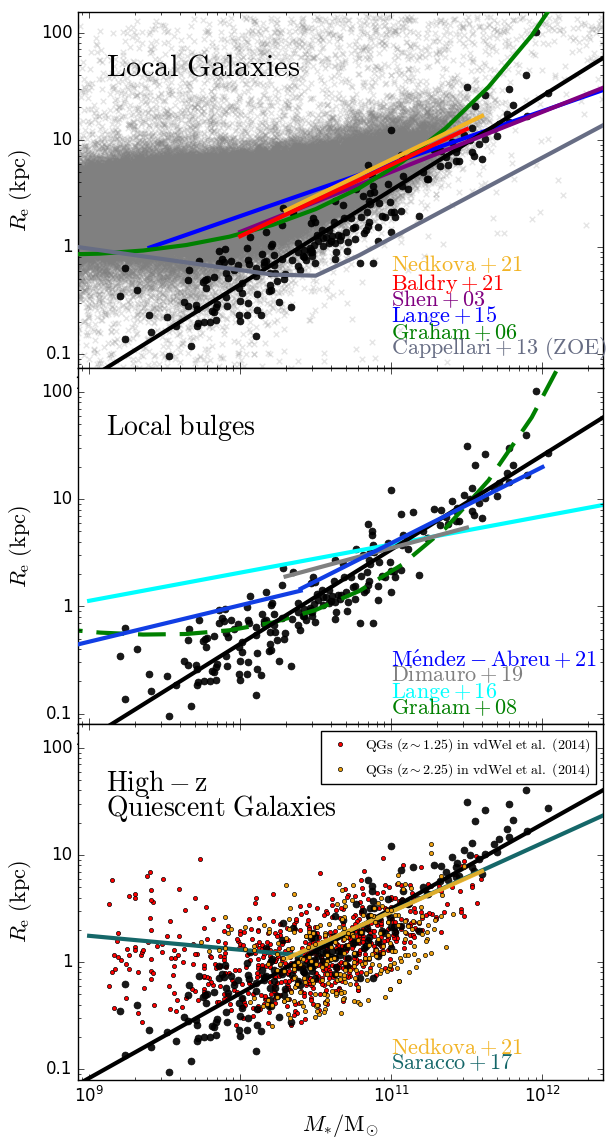}
    \caption{Comparison with the size-mass relations in the literature. 
    The solid black line is our single-power law fit to the {\em spheroid} data, and the black points are the spheroids presented  in Fig.~\ref{fig:size_mass_morphology}. 
    Upper panel: ETG, i.e., not just spheroid, size-mass relation at $z<1.0$ from \citet[][]{2003MNRAS.343..978S}, \citet[][$z<0.04$]{Graham2006A}, the galaxy ZOE from \citet[][$z\lesssim 0.01$]{Cappellari2013ATLAS}, \citet[][$0.002<z< 0.06$]{Lange2015}, \citet[][$0.04< z <0.15$]{Baldry2021}, and \citet[][$0.2<z<0.5$]{Nedkova2021}.
    The grey cloud depicts SDSS galaxies of all type.
    Middle panel: local bulge size-mass relation from \citet[][]{Graham2008}, \citet[][$0.002<z< 0.06$]{Lange2016}, \citet[][$z\sim0.25$]{Dimauro2019}, and \citet[][]{Mendez-Abreu2021}. 
    Bottom panel: the size--mass distribution for quiescent galaxies (QGs) at high-$z$.
    The red and orange points depict the QGs from \citet{van_der_Wel2014} at $z\sim1.25$ and $z\sim2.25$, respectively, with their effective radius scaled to equivalent axis (see texts).
    \citet{Nedkova2021}'s size-mass relation at $1.0<z<1.5$ and \citet{Saracco2017}'s relation ($1.2<z<1.4$) are shown in solid orange and dark green lines, respectively.}
\label{fig:size_mass_fit_coparison}
\end{figure}

\subsection{Comparison with early works on spheroid structural parameters}
\label{sec:early_work}

In this section, we compare the spheroid's structural parameters with some early pioneer works involving $R^{1/n}$ bulges \citep[][]{Khosroshahi2000A,Mollenhoff2001,Barway2009,Laurikainen2010}.
This will test if our spheroid parameters from multicomponent decomposition behave similarly compared to previous studies.

\citet[][]{Khosroshahi2000A} studied a sample of 26 early-type spiral galaxies from the UGC catalogue \citep[][]{1973ugcg.book.....N},  selected by \citet[][]{Balcells1994}. 
They performed two-dimensional $R^{1/n}$-bulge+exponential disc decompositions on the $K$-band images from \citet[][]{Andredakis1995} to obtain the bulge parameters.
\citet[][]{Mollenhoff2001} have studied 40 galaxies without a strong bar and with low inclination in the Revised ShapleyAmes Catalog \citep[][]{Sandage1981} via 2-D bulge+disc decomposition in $J$-, $H$-, and $K$-band.
Their sample consists of early-type spiral galaxies ranging from Sa to Sc type.
Here, we focus on their $K$-band data.
\citet{Barway2009} also performed 2-D bulge+disc decompositions on 36 bright field S0 galaxies from \citet[][]{Barway2005}, and imaged in the $K$-band. 
Finally, \citet[][]{Laurikainen2010} studied the structure of 175 local galaxies using deep $Ks$-band images.
Their sample consists of 117 S0, 22 S0/a and 36 Sa galaxies. 
Unlike the previous three works, \citet[][]{Laurikainen2010} performed a multicomponent fit to obtain the bulge parameters.
More specifically, in addition to an $R^{1/n}$ bulge and an exponential-disc, they included a Ferrer-bar when a bar was present.

Fig.~\ref{fig:3plot_compare} shows our spheroid parameters in contrast with the bulges' ones from the works above.
To compare bulges from different filters with our sample, we artificially shifted\footnote{Note that this is for illustrative purposes only. This action does not affect the slope of the scaling relation in the $\mu_\mathrm{0,Sph}$--$n_\mathrm{Sph}$ and $\mu_\mathrm{0,Sph}$--$\log(R_\mathrm{e,Sph})$ planes.} the central surface brightness, $\mu_\mathrm{0,Sph}$, of the bulges in \citet[][]{Khosroshahi2000B}, \citet[][]{Mollenhoff2001}, \citet[][]{Barway2009}, and \citet[][]{Laurikainen2010} by $\Delta\mu_\mathrm{0,Sph} \sim 2.48~\rm mag~arcsec^{-2}$.
That is, we applied a shift of $\mu_{i, \rm AB}-\mu_{K, \rm Vega} = 2.48~\rm mag~arcsec^{-2}$.
Our distribution overlaps quite well with the early works, albeit with a different degree of scatter.
Particularly in the $\log(R_\mathrm{e,Sph})$--$\log(n_\mathrm{Sph})$ and $\mu_\mathrm{0,Sph}$--$n_\mathrm{Sph}$ planes, the distributions follow a similar scaling relation.

While the overall shape of our distribution agrees well with the literature, discrepancies can be found in individual galaxies.
\citet[][]{Barway2009} bulges from S0 galaxies predominantly occupy the range of $3\lesssim n_\mathrm{Sph} \lesssim 4$ and the remainder of the sample scatter across $1 \lesssim n_\mathrm{Sph} \lesssim 3$.
Our S0 spheroids, however, mostly reside within $1.5\lesssim n_\mathrm{Sph} \lesssim 3$, a slightly lower value range compared to \citet[][]{Barway2009}.
Interestingly, some bulges from early-type S galaxy bulges in \citet[][green $\star$]{Khosroshahi2000B} and \citet[][grey $\blacksquare$]{Mollenhoff2001} have rather high S\'ersic indices ($n_\mathrm{Sph} \gtrsim 3.5$).
This is not the case for our spheroids and the bulges from \citet[][]{Laurikainen2010}.
Spheroids from our multicomponent analysis have a low S\'ersic index ($n_\mathrm{Sph} \lesssim 3.5$).
We speculate that during the fitting process, the S\'ersic function might be biased by the presence of a bar or an anti-truncated disc.

In Fig.\ref{fig:Re_Mag_fit}, we additionally compare our spheroid $\mathfrak{M}_\mathrm{Sph}$--$\log(R_\mathrm{e,Sph})$ data with those in the literature. 
Similar to the plots involving $\mu_0$, the absolute magnitude of the literature data is shifted by $\Delta \mathfrak{M}_\mathrm{Sph}\sim \rm 2.48~mag$ to match our spheroids. 
The solid black line is an ordinary least square (OLS) fit on our data points with a linear function.
It is not our intention to conduct a comprehensive meta-analysis of the literature but merely to demonstrate the similarity of some features between these distributions.
One can see that, collectively, the size--luminosity distribution across the four works aligns quite well with our spheroid relation, with an apparent point of distinction at low masses between some studies.
While \citet{Mollenhoff2001} and \citet{Laurikainen2010} observe a flattening of the bulge size--mass relation at the low- mass end, \citet{Barway2009} and \citet[][]{Khosroshahi2000A} do not.
However, some clarification is required.
As we can see in \citet[][their Fig.~13]{Gadotti2009}, the flattening at the faint-end occurs at $\log(M_\mathrm{*,Sph}/\rm M_{\odot}) \lesssim 9.5$ and only become evident when faint bulges are sampled.
The jump from the bulge sequence to the elliptical galaxy sequence is because of how the disc stars from S0+S0 galaxy merger add to the size and mass of the elliptical galaxy (Graham \& Sahu 2022, MNRAS, in Press).

\subsection{Distinction between giant elliptical and disc-galaxy}
\label{sec:E_S0_Sdistinction}

One can observe a significant overlap among massive S0 and S galaxy bulges---not to be confused with bars and inner discs---in all three structural relations (Fig.~\ref{fig:M_n_mu}). 
As one can see in the spheroid size-mass relation (Fig.~\ref{fig:size_mass_morphology}), the spheroids embedded in S0 and S galaxies are structurally similar.
To illustrate this point, we performed a symmetrical regression to the morphology-dependent size--mass relation in the left-hand panel of Fig.~\ref{fig:size_mass_morphology}.
Their bisector lines are:
\begin{subequations} 
\label{eq:size_mass_twofit}
\begin{align*}
 E+ES:
 \log(R_\mathrm{e,Sph}/\rm kpc)&= 0.96\log(M_\mathrm{*,Sph}/\rm M_{\odot})-9.94, \tag{18a}\\
 S0: 
 \log(R_\mathrm{e,Sph}/\rm kpc)&= 0.84\log(M_\mathrm{*,Sph}/\rm M_{\odot})-8.81, \tag{18b}\\
 S: 
 \log(R_\mathrm{e,Sph}/\rm kpc)&= 0.79\log(M_\mathrm{*,Sph}/\rm M_{\odot})-8.20, \tag{18c}\\
\end{align*}
\end{subequations}
with a scatter of $\Delta_{rms} = 0.18, 0.22$ and $0.23~\rm dex$ in the $\log(R_\mathrm{e,Sph})$ direction, respectively. 
Indeed, the slope, $S$, and intercept, $\rm int.$, for the bulges in S0 and S galaxies are very similar. 
Interestingly, the slope of the spheroid size--mass relation seem to be flattening\footnote{Note that the flattening disappears once we removed several outliers: the E galaxy with $\log(R_\mathrm{e, Sph}) >1.7$ and the two S galaxies with $\log(M_\mathrm{*, Sph}) < 9.4$. The resulting bisector lines for each group becomes: $\log(R_\mathrm{e,Sph}) = 0.87\log(M_\mathrm{*, Sph})-9.00$ for E galaxies,  
$\log(R_\mathrm{e,Sph}) = 0.84\log(M_\mathrm{*, Sph})-8.80$ for S0 galaxies, and 
$\log(R_\mathrm{e,Sph}) = 0.85\log(M_\mathrm{*, Sph})-8.80$ for S galaxies.} as we move from E to S galaxies.
A similar trend has been reported in \citet[][see their Fig.~13]{Gadotti2009} as the slopes of their size--mass relations decrease from E galaxies to classical bulges, and from classical to pseudo-bulges.
Although, their size--mass relations have considerably shallower slopes (by a factor of$\sim3$) than ours, with $S=0.38$ for E galaxies and $S=0.30$ for classical bulges \citep[see Section~4.3 in][]{Gadotti2009}.

The similarity between S0 and S bulges implies that they might share the same origin.
Indeed, the shared origin has previously been hinted at from different perspectives.
In stellar population studies \citep[][]{MacArthur2009}, bulges in both S0 and early-type S galaxies contain significantly older stars than the disc, implying the galaxy grows in an inside-out manner, a.k.a. disc-cloaking (H+22) upon an existing spheroid \cite[][]{Graham2015}.
From dynamical studies, some lower-mass S0 galaxies are suggested to be S galaxies with their spiral arms faded away \citep[see also][]{Aragon-Salamanca2006, Laurikainen2010}.
\citet[][]{Rizzo2018} investigated the disc dynamics of ten S0 galaxies in the CALIFA survey and found that in the specific angular momentum ($j_*$) versus stellar mass ($M_*$) diagram, their S0 discs align with the S discs from \citet[][see their Fig.~8]{Romanowsky2012}. 
This observation implies that some S0 galaxies are similar to massive spiral galaxies.
Some S galaxies may have evolved by slowly removing cold gas in their spiral arms via processes such as star formation \citep[][]{van_den_Bergh2009, Laurikainen2010, Williams2010, Cappellari2011b, Bellstedt2017} and outflow.
There is also significant overlap in the bulge-to-total ($B/T$) flux ratio of S galaxies and S0 galaxies \citep[e.g.,][H+22]{Mendez-Abreu2017}, giving credence to the fading spiral arm scenario.

\subsection{Modes of galaxy size evolution}
\label{sec:comparison}

\subsubsection{Comparison with local ETGs}
A log-linear size--mass relation was reported long ago in studies that used isophotal radii to describe galaxy size \citep[][]{Heidmann1969,Holmberg1969,Oemler1976,Strom1978}. 
As explained, for perhaps the first time in \citet[][]{Graham2019PASA}, this is understood in terms of the (stellar mass)--(S\'ersic $n$) relation for ETGs. 
The more slowly declining (with increasing radius) light profile of higher-$n$ ETGs results in larger isophotal radii being reached in ETGs with higher values of $n$.

Notably, our spheroid size-mass relation deviates from the \textit{curved} size-mass relation for ETGs and bulges.
Using the ETGs and dwarf ETGs (dETGs) imaged in the $B$-band from the compilation in \citet{Graham2003B}, they obtained two linear empirical scaling relations, $\mathfrak{M}$--$\log(n)$ and $\mathfrak{M}$--$\mu_0$.
With these two relations, \citet{Graham2006A} proceed to construct a \textit{curved} size--mass relations for ETGs (solid green line in the right-hand panel of our Fig.~\ref{fig:size_mass_morphology}).
Here, the ETG magnitudes are converted into stellar masses with a constant mass-to-light ratio\footnote{To ensure a fair comparison with the ETGs in the $B$-band from \citet{Graham2006A}, we use the same $M/L$ ratio prescription as was done in H+22 to convert ETG magnitudes into stellar mass.
\citet[][H+22's $M/L$ prescription]{Roediger2015} provided a colour-dependent $M/L$ relation in the $g$-band: $\log(M/L_g) = 1.379(g-i)-1.067$. 
Given that the SDSS $g$-band is similar to the $B$-band, we can approximate $M/L_B\approx M/L_g$.
We further assume ETGs have a constant colour of $(g-i)=1.2$ \citep{Fukugita1996}. 
This result in $M/L_B\approx3.9$.} of $M/L_B=3.9$.
Similar curved $\log(R_\mathrm{e})$--$\log(M_{*})$ distributions for galaxies have been found in other works \citep[e.g.,][]{2003MNRAS.343..978S, Lange2015, Nedkova2021}, in which a double-power law was used to fit the curved size-mass relation.
Similarly, \citet{Graham2008} have also produced a \textit{curved} size-mass relations for \textit{bulges}.
They used published decompositions for $\sim400$ S0 and S galaxies with predominantly two components: a S\'ersic-bulge plus an exponential-disc.
The scaling relations for these bulges were subsequently presented in \citet{Graham2019PASA}.
To compare the bulges data (taken in the $K$-band) with the ETGs data (taken in the $B$-band) in the $B$-band, we shifted the bulge magnitudes by $B_\mathrm{Vega}-K_\mathrm{Vega}=4.0~\rm mag$ and calculated their stellar masses with $M/L_B=3.9$.
The resulting \textit{curved} size-mass relation for bulges has a greater curvature than the galaxies' relation and leans into the more compact region in the diagram (dashed green line in the right-hand panel of Fig.~\ref{fig:size_mass_morphology}).
However, having data from more detailed decomposition (SG16, D+19, S+19, and H+22), we observe that spheroids follow a different relation which continues downwards to smaller sizes.
The tight spheroid size--mass relation shows that the size of a spheroid is a great predictor of stellar mass.
It also implies local spheroids in this mass range follow a simple scalar virial relation: $R_\mathrm{e} = (\beta~\mathrm{G} /\sigma_\mathrm{e}^{\alpha}) (M_{*}/\rm M_{\odot})^{\gamma}$ where G is the gravitational constant and $\alpha$, $\beta$, and $\gamma$ are some constants.

We further compare our spheroids to the galaxy and bulge size--mass relations in the literature.
Fig.~\ref{fig:size_mass_fit_coparison} shows a comparison of our spheroids with local galaxies (upper panel), local bulges/spheroids (middle panel), and high-$z$ quiescent galaxies (bottom panel).
Our spheroid size-mass relation (solid black line) and data (black points) are plotted in all three panels.
%***
%The bulk of the high-$z$ data appears consistent with local bulges. 
%If these compact, high-$z$ systems were to acquire/accrete/grow a large-scale disc, they would appear consistent with today's S0 and S galaxies.
%This is supportive of, albeit not conclusive, that `disc-cloaking'(H+22) may explain the disappearance of the `red nuggets' \citep[e.g.,][]{Daddi2005,van_Dokkum2008,Damjanov2009} seen at high-$z$.
%***

In the upper panel of Fig.~\ref{fig:3plot_compare}, the \textit{galaxy} size-mass relation from the following works are shown: \citet[][]{2003MNRAS.343..978S}, \citet[][]{Graham2006A}, \citet[][]{Lange2015}, \citet[][]{Baldry2021}, \citet[][their quiescent galaxies (QGs) at $0.2<z<0.5$ ]{Nedkova2021}.
Additionally, the empirical zone of exclusion \citep[ZOE,][]{Bender1992,Burstein1997}, defined in \citet{Cappellari2013ATLAS}, is shown with a bent solid grey line.
The grey cloud depicts the SDSS galaxies from the NASA-Sloan ATLAS catalogue at $z<0.05$.

Our \textit{spheroid} relation resides on the high-mass and small-radius side of the SDSS galaxies.
It also presents a steeper slope ($S=0.88$) compared to the galaxy relations.
Simply put, the spheroids are more compact than ETGs or QGs.
At the high-mass end ($M_* > 4 \times10^{11}~\rm M_{\odot}$), the \textit{galaxy} size-mass relation from \citet{2003MNRAS.343..978S}, \citet{Lange2015},  \citet{Baldry2021}, and \citet[][$0.2<z<0.5$]{Nedkova2021} align better with our spheroid relation than compared to the low-mass end ($M_* < 4 \times10^{11}~\rm M_{\odot}$).
This is because single-component spheroidal E galaxies dominate the high-mass range.
Our spheroids at the high-mass end are elliptical galaxies modelled with a single S\'ersic function similar to the earlier works.
E galaxies appear to follow the same continuous log-linear relation as the embedded spheroids in S0 and S galaxies.
Indeed, \citet{Lange2016} show that their galaxy size-mass relation was the steepest ($S=0.786$) when they only considered massive galaxies ($M_* > 2\times10^{10}~\rm M_{\odot}$, see their Table~1), a result much closer to our relation.
Further, the ZOE presented in \citet{Cappellari2013ATLAS} is an empirical model of a double-power law that cuddles the lower bound of their galaxy size-mass distribution.
It depicts an empirical lower limit for galaxies' size and maximum density.
However, note that \citet{Cappellari2013ATLAS}'s ZOE is constructed with local \textit{galaxies} (within $42~\rm Mpc$). 
The embedded \textit{spheroids} need not conform to such a constraint.
If indeed some spheroids are already fully formed at high-$z$, the ZOE at higher $z$ could be different to the ZOE in \citet{Cappellari2013ATLAS}.
Since the spheroids' relation roughly follows a log-linear trend across $2\times10^{9}\lesssim M_*/\rm M_{\odot}\lesssim 2\times10^{12}$, its ZOE should also be log-linear.

\subsubsection{Comparison with local bulges}

Our result also highlights the necessity for multicomponent decomposition in order to extract bulges correctly.
In the middle panel of Fig.~\ref{fig:size_mass_fit_coparison}, we compare our spheroids with some \textit{bulge} size-mass relations at low-$z$: \citet[][]{Graham2008}, \citet[][]{Lange2016}, \citet[][]{Dimauro2019}, and \citet[][]{Mendez-Abreu2021}.
Building on \citet{Allen2006}, \citet{Lange2016} performed an automatic Bulge+Disc decomposition of 2247 ETGs and LTGs with stellar mass $10^{9}\lesssim M_*/\rm M_{\odot}<3\times10^{11}$ at $0.002<z<0.06$ taken from the Galaxy And Mass Assembly (GAMA) survey \citep[GAMA II,][]{Driver2009,Driver2016, Liske2015}. 
They improved on the conventional fitting routines by performing multiple fits with a wide range of starting parameters and using the median value of these fits to obtain the size--mass relation.
However, their bulge size-mass relations differ significantly from our spheroid relation, with the slope $S=0.263$ \citep[][see their Table~1]{Lange2016}.
In \citet{Dimauro2019}, the slope of the size--mass relation iss $S=0.385$ for all bulges coming from galaxies with $M_\mathrm{*, Sph} > 2\times 10^{10}$ at $z\sim0.25$.
Since both their methods do not account for potentially biasing components, e.g., bars, disc truncation and anti-trucation, and nuclear components, the bulge sizes are likely overestimated.
On the contrary, \citet[][]{Mendez-Abreu2017,Mendez-Abreu2021} present a bulge relation made using 2D multi-component decomposition for 404 galaxies at $z<1$ in the Calar Alto Legacy Integral Field Area survey \citep[CALIFA-DR3,][]{Garcia-Benito2015, Sanchez2016} that includes a variety of components, specifically, the bulge, bar, nuclear point source, and Type I-III disc.
As a result, their bulge size--mass relation is the closest to ours, with a slope of $S=0.71$ for bulges with $10.5 < \log(M_\mathrm{*,Sph}/\rm M_{\odot}) < 12$. 
Interestingly, however, the low-mass ($8.0 < \log(M_\mathrm{*,Sph}/\rm M_{\odot} < 10.5$) bulges in \citet{Mendez-Abreu2021} follows a shallower relation with a slope of $S=0.34$, resulting in an `up-bend' in the overall size--mass relation.
Similar up-bend at $M_\mathrm{*,Sph}\sim10^{8}$--$3\times10^{9}~\rm M_{\odot}$ was also reported in other works \citep[e.g.,][]{Gadotti2008,Laurikainen2010}.

We note that the absent of up-bend in our relation could be a result of sample selection.
From Fig.~8 in \citet[][]{Laurikainen2010}, we can see that bulges from Sc or later-type spiral (Sc-) galaxies dominate the low-mass range ($M_\mathrm{*,Sph}\sim10^{8}$--$3\times10^{9}~\rm M_{\odot}$).
Without the Sc- bulges, their spheroid size--luminosity distribution resembles a log-linear relation.
Sc- galaxies are also abundant in the parent sample
of the aforementioned work that presented the up-bend \citep[][]{Gadotti2009, Lange2016, Mendez-Abreu2021}. 
However, since our data sources, H+22 and D+19, contain mostly early-type (Sa-Sb) spiral galaxies, the up-bend is not present in our spheroid size--mass relation.
While our sample does not contain many late-type (Sc-Sd) spiral galaxies, or more specifically, galaxies with $M_{\rm *,Sph} < 2\times10^9~\rm M_\odot$, the light-green curve in Fig.~\ref{fig:size_mass_morphology} suggests a possible flattening of the size-mass relation at low masses. 
Late-type spiral galaxies can, however, be challenging to model because they typically contain bulges whose surface brightness is relatively faint compared to the inner disc \citep[][their Fig.~21]{Graham2001A}, and with ground-bases seeing, the bars of bulgeless spiral galaxies can mimic bulges \citep[e.g.][]{2015ApJ...809L..14B, 2017ApJ...850..196B}. 
While our colleagues have addressed these issues and measured a reduction of the slope in the size-mass diagram for late-type spiral galaxies, we do not attempt to do this.

When the disc (and any additional inner-disc) and disc-induced components (e.g., bars and spiral arms) are removed, the remaining component would be the relatively dense spheroid.
Galaxy size-mass relations possess a flatter slope than spheroids' because the disc in S0 and S galaxies, which exist at lower-mass end, increases the overall size and decreases the overall density of a galaxy.
The deviation between the galaxy and spheroid size-mass relation is more prominent at the low-mass end ($M_\mathrm{*, gal}/\rm M_{\odot} \lesssim 10^{11}$), where the galaxies tend to have a lower bulge-to-total flux ratio and the disc's mass trumps the spheroids'. 

\subsubsection{Comparison with high-$z$ quiescent galaxies}

With the knowledge that ETGs have a different size-mass relation than the spheroids, kicking up in size at low-mass end because of the disc, we compare our spheroid relation with the quiescent galaxies at $z>1.0$.
The bottom panel of Fig.~\ref{fig:size_mass_fit_coparison} shows the galaxy size-mass relation at $z>1.0$: \citet[][QGs at $1.2<z<1.4$]{Nedkova2021} and \citet[][QGs at $1.2<z<1.4$]{Saracco2017}.
The red and orange points are the QGs from CANDELS \citep[][]{van_der_Wel2014} at $z\sim1.25$ and $z\sim2.25$ (their Fig.~5)\footnote{The original size-mass distribution was depicted using the effective radius in the major-axis instead of the circularised radius. Following what was done in \citet[][]{Saracco2017}, we scaled down their radius by the average radius ratio $\langle R_\mathrm{e,circularised}/R_\mathrm{e,major} \rangle\sim 0.76$ \citep[][]{Cappellari2013size-mass} to match the size-mass relations in the other studies.}, respectively.
The relation in \citet{Nedkova2021} matches the high-mass ($M_\mathrm{*,gal} / \rm M_{\odot} \gtrsim 3\times10^{10}$) end of the relation in \citet{Saracco2017} remarkably well. 
Compared to the \citet{Nedkova2021} galaxy size-mass relation at $0.2<z<0.5$ (see the upper panel of Fig.~\ref{fig:size_mass_fit_coparison}), their relation at $z>1.0$ migrates toward the lower-right side of the plot, implying quiescent galaxies are more compact as redshift increases \citep[see also][]{van_der_Wel2014}.
The up-bend at $M_*\sim3\times10^{10}~\rm M_{\odot}$ in the \citet{Saracco2017} galaxy size-mass relation is also present in the QGs from \citet{van_der_Wel2014} at $z\sim1.25$--$2.25$.
The QGs from \citet{van_der_Wel2014} at $z\sim1.25$--$2.25$ follow a similar trend to our spheroids' size-mass relation. 
In between $10^{10} \lesssim M_*/\rm M_{\odot} \lesssim10^{11}$, our spheroids reside in the middle of the \citet{van_der_Wel2014}'s QGs distribution. 
It shows that, in terms of size and mass, local spheroids are structurally similar to high-$z$ quiescent galaxies. 
The galaxy size evolution among QGs has been discussed by many studies \citep[e.g.,][]{Trujillo2007, Bezanson2009, Taylor2010, Barro2013,van_der_Wel2014,van_Dokkum2015}, where QGs become less compact as $z$ decreases.
Since local spheroids are analogous to the quiescent system at $z>1.0$, our result supports the scenario where the system builds from the inside-out, either through the disc-cloaking process \citep[][]{Graham2015, Hon2022} to build up a disc within over an existing spheroid to become S0 or S galaxy \citep[see also][]{Costantin2020, Costantin2022}, or major mergers to become an elliptical galaxy (Graham \& Sahu 2022, in Press).

\section{Summary} 
\label{sec:summary}
In this paper, using the local bulge/spheroid data from multi-component decompositions in SG16, D+19, S+19, and H+22, we present the followings:

\begin{itemize}
    \item
    The distribution of the spheroids'  $i-$band absolute magnitude ($\mathfrak{M}_\mathrm{Sph}$) against their structural parameters, namely the spheroid's S\'ersic index ($n_\mathrm{Sph}$), central surface brightness ($\mu_\mathrm{0,Sph}$), and effective radius ($R_\mathrm{e,Sph}$) have been presented in Section \ref{sec:Struc_para}.

    \item
    The correlation strength among $R_\mathrm{e,Sph}$, $\mu_\mathrm{0,Sph}$, and $n_\mathrm{Sph}$ are measured (Section~\ref{sec:Corr_among_para}).
    
    \item
    The $\mathfrak{M}_\mathrm{Sph}$--$\log(R_{z,\rm Sph})$ and  $\mathfrak{M}_\mathrm{Sph}$--$\mu_{z,\rm Sph}$ relations, using the different scale radii $R_\mathrm{z, \rm Sph}$, enclosing different fractions of the spheroid light, are presented in Section~\ref{sec:R_variation}.

    \item
    The spheroid mass ($M_\mathrm{*,Sph}$) versus S\'ersic index ($n_\mathrm{Sph}$), projected mass density ($\Sigma_\mathrm{0,Sph}$), and effective radius ($R_\mathrm{e,Sph}$) relations are provided in Section~\ref{sec:sigma0}.
    Their behaviour resembles that of the $\mathfrak{M}_\mathrm{Sph}$--$\log(n_\mathrm{Sph})$, $\mathfrak{M}_\mathrm{Sph}$--$\mu_\mathrm{0,Sph}$ and $\mathfrak{M}_\mathrm{Sph}$--$\log(R_\mathrm{e,Sph})$ relations, with similar correlation strength.

    \item 
    We charted the correlation strength ($r_p, r_s$) as a function of the fraction of light, $z$, included within the scale radii $R_\mathrm{z, \rm Sph}$ for different parameter ($R_{z, \rm Sph}$, $\Sigma_{z, \rm Sph}$,  $\langle\Sigma\rangle_{z, \rm Sph}$, and $B_{z, \rm Sph}(n_\mathrm{Sph})$) pairings with the spheroid mass (Section~\ref{sec:corr_trend}).
    Among them, the  $\log(M_\mathrm{*,Sph})$--$\log(R_{z,\rm Sph})$ relations consistently have the highest correlation strength ($r_p, r_s \sim 0.9$) across all $z$.
    The S\'ersic `shape function' $\log(M_\mathrm{*,Sph})$--$\log(B_\mathrm{z, \rm Sph}(n_\mathrm{Sph})$ relations have the second highest correlation across all $z$, with $r_p$ and $r_s \sim 0.7$, while the strength of the $\log(M_\mathrm{*,Sph})$--$\log(\Sigma_{z, \rm Sph})$ and $\log(M_\mathrm{*,Sph})$--$\log(\langle \Sigma \rangle{z, \rm Sph})$ relations vary significantly with the choice of $z$.

      \item 
    The local spheroid size ($R_\mathrm{e,Sph}$)--mass ($M_\mathrm{*,Sph}$) relation is presented in Section~\ref{sec:size_mass}.
    For the full sample, the bisector regression line is: 
   \begin{equation}
       \log(R_\mathrm{e, Sph}/\rm kpc) = 0.88 \log(M_\mathrm{*, Sph}/\rm M_{\odot})-9.15,
   \end{equation} 
   with an intrinsic scatter of $\Delta_{rms} = 0.24~ \rm dex$.   
   \item
   Four additional scaling relations: $\log(n_\mathrm{Sph})$--$\log(M_\mathrm{*,Sph})$, $\log(B_\mathrm{e,Sph})$--$\log(M_\mathrm{*,Sph})$, $\log(\Sigma_\mathrm{0,Sph})$--$n_\mathrm{Sph}$, and $\log(R_\mathrm{e,Sph})$--$\log(n_\mathrm{Sph})$ are presented in Section~\ref{sec:additional_relations}.
   We obtained the bisector lines:
    \begin{equation}
       \log(n_\mathrm{Sph}) = 0.43~ \log(M_\mathrm{Sph}/\rm M_{\odot})-4.20,
   \end{equation} 
   with an intrinsic scatter of $\Delta_{rms} = 0.21~ \rm dex$; 
    \begin{equation}
       \log(B_\mathrm{e, Sph}) = 0.20~ \log(M_{\rm *,Sph}/\rm M_{\odot}) - 1.70,
   \end{equation} 
   with an intrinsic scatter of $\Delta_{rms} = 0.10~ \rm dex$; 
    \begin{equation}
       \log(\Sigma_\mathrm{0, Sph}/\rm M_{\odot}pc^{-2}) = 0.59~ n_\mathrm{Sph}+3.42,
   \end{equation} 
   with an intrinsic scatter of $\Delta_{rms} = 0.47~ \rm dex$; and 
    \begin{equation}
       \log(R_\mathrm{e, Sph}/\rm kpc) = 2.06~ n_\mathrm{Sph} - 0.66,
   \end{equation} 
   with an intrinsic scatter of $\Delta_{rms} = 0.39~ \rm dex$. 
   In each case, the scatter is measured in the vertical direction.
\end{itemize}

Here, we briefly summarise the findings in this paper:
\begin{enumerate}

\item There is no clear boundary between spheroids embedded in S0 and S galaxies. 
They have a significant overlap in the size-mass diagram and other parameter planes, indicating a shared origin or, at the very least, governing formation physics among bulges (see Section~\ref{sec:E_S0_Sdistinction}).

\item The spheroid radius ($R_{z, \rm Sph}$) is the best predictor of its luminosity and stellar mass, regardless of the fraction of light, $z$, used.
The shape of the spheroid's light profile ($n_\mathrm{Sph}$) and `shape function', $B_{z, \rm Sph}(n_\mathrm{Sph})$, also correlate well with the spheroid mass, albeit it not as strong as $R_{z, \rm Sph}$.
The surface brightness, $\mu_{z, \rm Sph}$, and the surface mass densisities, $\Sigma_{z, \rm Sph}$ and $\langle\Sigma\rangle_{z, \rm Sph}$, are weak predictors of the spheroid stellar mass because of their low correlation with $M_\mathrm{*, Sph}$, a result which is also subject to the light fraction $z$.

\item  The spheroid size ($R_\mathrm{e,Sph}$)--mass ($M_\mathrm{*,Sph}$) distribution exhibits a roughly log-linear relation at $M_\mathrm{*,Sph} \gtrsim 10^{9}~ \rm M_{\odot}$, contrary to the curved size--mass predicted in \citet{Graham2006A} for ETGs and the double-power law from other works.
Since the disc (and other) components are less dense than the bulge, S0 and S galaxies will have a larger size than a pure spheroid at the same mass, resulting in the upbend in the low-mass end where discs are more prominent.

\item The spheroid central surface brightness ($\mu_\mathrm{0,Sph}$)--S\'ersic index ($n_\mathrm{*,Sph}$) relation shows a strong \textit{linear} trend (Fig.\ref{fig:L_plot}).
As such, in log-space ($\mu_\mathrm{0,Sph}$--$\log(n_\mathrm{*,Sph})$), the relation appears curved.
This is in contrast to the linear scaling $\mu_\mathrm{0,Sph}$--$\log(n_\mathrm{*,Sph})$ relation presented in \citet[][]{Khosroshahi2000B}.
If the `photometric plane' is present in our sample, we speculate its surface will also be curved if using $\log(n)$.

\item  Our spheroid size-mass relation is a factor of$\sim 3$ steeper than some reported bulge size-mass relation obtained via Bulge+Disc decomposition.
Likely due to the lack of Sc or later-type spiral galaxies in our sample, we do not see the flattening of slope (i.e. an up-bend) in the low-mass end ($M_\mathrm{*,Sph}/\rm M_{\odot} < 10^{9}$) that is common in other bulge size--mass relation in the literature. 
Finally, our spheroids' relation aligns well with the quiescent galaxy size-mass relation at $z\sim1.25$ and $2.25$ in between $2\times10^{10}<M_{*}/\rm M_{\odot} <4\times10^{11}$.
It indicates that local spheroids are structurally similar to the high-$z$ quiescent galaxies.

\end{enumerate}

\section*{Acknowledgements}
\addcontentsline{toc}{section}{Acknowledgements}
This research was supported under the Australian Research Council's funding scheme DP17012923. 
This research has made use of the NASA/IPAC Extragalactic Database (NED), which is funded by the National Aeronautics and Space Administration and operated by the California Institute of Technology.

The project is made possible by using the following software packages:
\texttt{AstroPy} \citep{Astropy2013, Astropy2018},
\texttt{Cmasher} \citep{van_der_Velden2020},
\texttt{IRAF} \citep{Tody1986IRAF, Tody1993IRAF},
\texttt{ISOFIT} \citep{Ciambur2015},
\texttt{Matplotlib} \citep{Hunter:2007},
\texttt{NumPy} \citep{Harris2020array_NumPy},
\texttt{Profiler} \citep{Ciambur2016},
\texttt{SAOImageDS9} \citep{Joye2003DS9},
\texttt{SciPy} \citep{2020SciPy},
\texttt{SExtractor} \citep{Bertin1996}, and
\texttt{TOPCAT} \citep{Taylor2005Topcat}.

All the scripts used in the analysis are available on GitHub (\url{https://github.com/dex-hon-sci/GalSpheroids}).

%%%%%%%%%%%%%%%%%%%%%%%%%%%%%%%%%%%%%%%%%%%%%%%%%%
\section*{Data Availability}
\addcontentsline{toc}{section}{Data Availability}

The data underlying this article will be shared upon request to the corresponding author.



\bibliographystyle{mnras}
\bibliography{paperII_bib} % if your bibtex file is called example.bib

%\begin{thebibliography}{99}
%\end{thebibliography}

%%%%%%%%%%%%%%%%%%%%%%%%%%%%%%%%%%%%%%%%%%%%%%%%%%

%%%%%%%%%%%%%%%%% APPENDICES %%%%%%%%%%%%%%%%%%%%%

%\appendix


%%%%%%%%%%%%%%%%%%%%%%%%%%%%%%%%%%%%%%%%%%%%%%%%%%

% Don't change these lines
\bsp	% typesetting comment
\label{lastpage}
\end{document}